\documentclass[%
 reprint,
 superscriptaddress,
 amsmath,
 amssymb,
 aps,
 prab,
 floatfix,
]{revtex4-2}

\usepackage[english]{babel}
\usepackage{textcomp}
\usepackage[T1]{fontenc}
\usepackage{lmodern}
\usepackage{tgtermes}
\usepackage{newtxmath}
\usepackage{siunitx}
\usepackage{physics}
\usepackage{graphicx}
\usepackage{dcolumn}
\usepackage{bm}
\usepackage[hidelinks,breaklinks=true]{hyperref}
\usepackage{subcaption}
\DeclareCaptionJustification{revtexjustified}{\flushing}
\usepackage[most]{tcolorbox}
\usepackage{tikz}
\usetikzlibrary{calc,arrows.meta}
\usetikzlibrary{arrows.meta}

\DeclareRobustCommand{\IPPL}{IPPL}
\DeclareRobustCommand{\OPAL}{OPAL}
\DeclareRobustCommand{\OPALX}{OPALX}
\DeclareRobustCommand{\Kokkos}{Kokkos}
\DeclareRobustCommand{\orbitt}{\texttt{OrbitThreader}}
\DeclareRobustCommand{\gfbl}{G4Beamline}
\DeclareRobustCommand{\trw}{TraceWin}

\begin{document}

\preprint{APS/123-QED}

\title{\OPALX: A performance-portable framework for particle accelerator simulations}

\author{M.~Zampetakis}
\affiliation{Paul Scherrer Institute, 5232 Villigen PSI, Switzerland}

\author{R.~S.~Ammann}
\affiliation{Paul Scherrer Institute, 5232 Villigen PSI, Switzerland}
\affiliation{Department of Physics, ETH Zurich, 8093 Zurich, Switzerland}

\author{Y.~Bi}
\affiliation{School of Science, Shenzhen Campus of Sun Yat-sen University, Shenzhen 518107, China}

\author{J.~Biddiscombe}
\affiliation{Swiss National Supercomputing Centre (CSCS), ETH Zurich, 6900 Lugano, Switzerland}

\author{P.~Calvo}
\affiliation{Centro de Investigaciones Energ\'eticas, Medioambientales y Tecnol\'ogicas (CIEMAT), 28040 Madrid, Spain}
\affiliation{Carlos III University of Madrid (UC3M), Department of Physics, 28911 Leganés, Spain}


\author{C.~Jolly}
\affiliation{STFC Rutherford Appleton Laboratory, Harwell Campus, Oxfordshire OX11 0QX, United Kingdom}

\author{S.~A.~Klapproth}
\affiliation{Paul Scherrer Institute, 5232 Villigen PSI, Switzerland}
\affiliation{Department of Physics, ETH Zurich, 8093 Zurich, Switzerland}

\author{A.~P.~Liemen}
\affiliation{Paul Scherrer Institute, 5232 Villigen PSI, Switzerland}
\affiliation{Department of Physics, ETH Zurich, 8093 Zurich, Switzerland}

\author{S.~Muralikrishnan}
\affiliation{J\"ulich Supercomputing Centre, Forschungszentrum J\"ulich, 52425 J\"ulich, Germany}


\author{C.~Rogers}
\affiliation{STFC Rutherford Appleton Laboratory, Harwell Campus, Oxfordshire OX11 0QX, United Kingdom}

\author{M.~Sadr}
\affiliation{Paul Scherrer Institute, 5232 Villigen PSI, Switzerland}
\affiliation{Massachusetts Institute of Technology, Cambridge, Massachusetts 02139, USA}


\author{J.~Thompson}
\affiliation{STFC Rutherford Appleton Laboratory, Harwell Campus, Oxfordshire OX11 0QX, United Kingdom}


\author{A.~Adelmann}
\email[Corresponding author: ]{andreas.adelmann@psi.ch}
\affiliation{Paul Scherrer Institute, 5232 Villigen PSI, Switzerland}

\date{\today}

\begin{abstract}
    Accurate charged-particle beam simulations require self-consistent collective fields and realistic accelerator models, while their computational cost motivates efficient use of heterogeneous computing systems. We present \OPALX, a performance-portable accelerator simulation framework that builds on the beam-dynamics capabilities of \OPAL. Its modular architecture uses the Independent Parallel Particle Layer~(\IPPL) and Kokkos to support distributed-memory execution on multicore CPUs and GPUs from a common codebase. The framework combines relativistic particle tracking with prescribed accelerator fields, particle-in-cell space charge calculations, and spin tracking. We describe the physical and numerical models and their integration into the simulation workflow. The implementation is assessed using disorder-induced heating and frozen-spin test cases, together with comparisons of spin evolution with \gfbl{} and beam transport with \trw. A three-dimensional space charge benchmark evaluates throughput on CPU and GPU systems, reaching approximately $10^9$ macroparticles per second on a single NVIDIA A100 GPU for this benchmark. These results provide initial verification of the implemented models and demonstrate CPU and GPU execution of self-consistent space charge calculations. OPALX provides a common foundation for extending accelerator models and numerical methods across heterogeneous computing platforms.
\end{abstract}


\maketitle

\section{Introduction} \label{sec:1}
Numerical simulations are essential to the design and operation of particle accelerators. For high-intensity and high-brightness beams, accurate predictions require realistic accelerator fields and a self-consistent treatment of collective effects. Space charge can modify beam sizes, emittances, halo formation, and particle losses, particularly in low-energy accelerators and injectors~\cite{Reiser:2008, Hofmann:2017}. In dense beams, discrete Coulomb interactions can introduce additional evolution beyond the smooth space charge field~\cite{Prat:2022, Mitchell:2015}. The interplay between space charge and intrabeam scattering~(IBS) can further affect beam evolution in regimes where both collective and collisional effects are significant~\cite{Zampetakis:2024, Zampetakis:2025}. Resolving these processes can require large particle ensembles, fine meshes, and many integration steps, making efficient parallel computation an integral part of accelerator modeling~\cite{Qiang:2000, Huebl:2022}.

The Object Oriented Parallel Accelerator Library~(\OPAL) provides an established foundation for these calculations~\cite{OPAL}. Its \texttt{OPAL-t} mode models beam lines, linacs, and radio-frequency~(RF) photo-injectors, while \texttt{OPAL-cycl} addresses cyclotrons and fixed-field alternating-gradient~(FFA) accelerators. Applications include high-charge electron-beamline optimization at the Argonne Wakefield Accelerator~\cite{Neveu:2019} and cyclotron injection and high-current cyclotron design~\cite{Winklehner:2017, Winklehner:2022}. These applications combine accelerator-component models with parallel particle tracking and self-consistent space charge calculations. The resulting algorithms, physical models, and operational experience form the basis for the development of \OPALX{}.

The increasing use of heterogeneous computing systems creates a need to execute these calculations efficiently on both multicore CPUs and GPUs. Maintaining separate implementations for each architecture would complicate the development and verification of accelerator-physics models. \OPALX{} addresses this challenge through a modular computational framework based on the Independent Parallel Particle Layer~(\IPPL)~\cite{IPPL} and \Kokkos{}~\cite{Kokkos}. \IPPL{} supplies distributed particle and field data structures, particle--mesh operations, and field solvers; \Kokkos{} provides abstractions for parallel execution and memory management within each process. Together with communication through the Message Passing Interface~(MPI)~\cite{MPI}, these dependencies allow accelerator models to use a common source implementation across CPU and GPU platforms.

The contribution of \OPALX{} is the integration of accelerator modeling and tracking within this computational framework. The program organizes particle advancement, accelerator components, external-field evaluation, collective-field calculations, and diagnostics through common interfaces. Established \OPAL{} capabilities can therefore be migrated and verified incrementally, while additional models can be introduced within the same workflow. The framework is intended for both hadron and lepton accelerators. The models discussed here include relativistic orbital dynamics, spin tracking, and space charge calculations using three-dimensional and 2.5D particle--mesh methods and a particle--particle particle--mesh~(P3M) method for short-range interactions. Beam generation, emission, and accelerator-element placement connect these models to the specification of an accelerator simulation. The source code and documentation are publicly available~\cite{OPALX, OPALXManual}.

This paper presents the computational framework, the physical and numerical models, and their integration into the simulation workflow. Its assessment focuses on the results available in the present version. Disorder-induced heating and frozen-spin test cases examine short-range interactions and spin dynamics, respectively. Comparisons with \gfbl{}~\cite{roberts_g4beamline_2007} and \trw{}~\cite{UriotTraceWin:2015} provide independent checks of spin evolution and beam transport. A three-dimensional space charge benchmark~\cite{Huebl:2026} measures throughput on CPU and GPU systems. These studies establish the behavior of selected models and their execution on different architectures; broader verification and accelerator applications remain part of the ongoing development.

This paper is organized as follows. Section~\ref{sec:design} introduces the design principles and computational framework of \OPALX{}. Section~\ref{sec:physics} presents the physical and numerical models, including particle and spin dynamics, time integration, the self-field formulation, particle--field coupling, and space charge solvers. Section~\ref{sec:architecture} describes the program structure and simulation workflow, including input processing, tracking, diagnostics, and reproducibility. Section~\ref{sec:components} discusses beam generation, accelerator-element placement, and external-field evaluation. Section~\ref{sec:validation} presents numerical verification cases and comparisons with independent simulation codes. Section~\ref{sec:performance} describes the performance benchmark methodology and CPU and GPU results. Section~\ref{sec:outlook} discusses current limitations and planned developments, and Sec.~\ref{sec:conclusions} summarizes the main conclusions.

\section{Design principles and computational framework} \label{sec:design}

\subsection{Design objectives}

High-fidelity accelerator simulations can require large particle ensembles, fine computational meshes, and tracking over many integration steps. Distributed-memory particle-in-cell~(PIC) methods have long enabled such simulations to use large-scale computing resources~\cite{Qiang:2000}. More recently, the increasing use of heterogeneous systems combining multicore CPUs and GPUs has created a need for accelerator codes that can use different hardware architectures without maintaining separate implementations of their physical models~\cite{Huebl:2022}.

A central design objective of \OPALX{} is therefore performance portability. In this context, performance portability denotes the ability to express computationally intensive algorithms through a common source implementation and execute them efficiently on different CPU and GPU systems. It does not imply identical performance on every architecture: particle tracking, particle--mesh operations, field solvers, and communication may place different demands on the available hardware. Instead, the objective is to retain a common implementation of the physical and numerical models while delegating architecture-specific execution and memory management to the underlying parallel framework.

A second objective is distributed-memory scalability. Simulations should be able to use multiple processes and compute nodes when the particle number or mesh resolution required for a calculation exceeds the capacity of a single device. The computational framework must therefore combine interprocess communication with the node-level parallelism available from multicore CPUs or GPUs. The resulting execution model is intended to support simulations ranging from local development and verification cases to large accelerator applications on high-performance computing systems.

\OPALX{} builds on the accelerator-physics experience accumulated in \OPAL{}~\cite{OPAL}. Established physical models, accelerator descriptions, and simulation concepts are retained or migrated where appropriate. However, \OPALX{} is not intended as a line-by-line port of the legacy code. Its computational structure has been redesigned to separate accelerator-physics functionality from parallel particle and field operations. This approach permits established \OPAL{} capabilities to be migrated and verified incrementally while providing a foundation for new physical models, accelerator components, and numerical methods.

\subsection{Layered computational framework}

The \OPALX{} computational framework is organized in layers with distinct responsibilities. At the application level, \OPALX{} provides the accelerator-specific functionality. This includes the description of the beam and accelerator lattice, the representation of accelerator components, the evaluation of external and collective fields, particle tracking, particle-loss treatment, and diagnostics. This layer defines the physical problem and coordinates the overall simulation.

\IPPL{}~\cite{IPPL} provides the distributed particle and field abstractions needed in \OPALX{}. It supplies the infrastructure required to represent particle ensembles and mesh-based fields across multiple processes, together with mechanisms for domain decomposition, particle redistribution, communication between subdomains, particle--field operations, and field solvers. Accelerator-physics models can therefore operate on particles and fields without directly implementing the associated distributed-memory data management and numerical methods.

\Kokkos{} provides abstractions for node-level parallel execution and memory management~\cite{Kokkos}. Computational kernels can therefore be executed using the selected CPU or GPU backend without embedding hardware-specific code in the accelerator-physics models. MPI~\cite{MPI} provides communication between processes on the same or different compute nodes. \Kokkos{} and MPI consequently address complementary levels of parallelism: \Kokkos{} exposes parallelism within a process, whereas MPI distributes the simulation across processes and nodes.

In a distributed simulation, the computational domain and its associated field data are partitioned among MPI processes. Each process operates primarily on its local particles and field data. Particles are redistributed as they move between subdomains, while field operations exchange information between neighboring subdomains where required. Global quantities needed by numerical algorithms or diagnostics are obtained using collective communication.

Together, \IPPL, \Kokkos, and MPI isolate much of the numerical, hardware-dependent and communication-related functionality from the accelerator-physics layer. The physical models implemented in \OPALX{} can consequently operate in terms of particles, fields and accelerator elements rather than numerical approaches related to particle--field couplings and field solvers or high-performance computing~(HPC) strategies such as device memory, thread scheduling, and explicit message-passing operations.

This separation allows novel numerical methods and HPC strategies to be implemented, tested, and optimized before they are used in complete accelerator simulations. Mini-applications in \IPPL~\cite{IPPL} support this workflow. It also enables alternative solver implementations to be selected without changing the general tracking workflow. Similarly, new accelerator components can be introduced through the common mechanisms for element placement and external-field evaluation.

\section{Physical and numerical models} \label{sec:physics}
\OPALX{} combines relativistic particle tracking with external accelerator fields, self-consistent collective effects, and optional spin and particle-decay processes. This section presents the physical equations, numerical approximations, and field solvers implemented in the framework.

\subsection{Particle equations of motion}
The beam is represented by an ensemble of macroparticles. Each macroparticle represents a group of physical particles of a given species. Its statistical weight determines its contribution to the collective charge distribution, while its motion in an electromagnetic field is governed by the charge-to-mass ratio of the corresponding particle species.

For macroparticle $i$ of species $s$, the orbital state is described by its position $\mathbf{r}_i=(x_i,y_i,z_i)$ in the Cartesian laboratory frame and its normalized mechanical momentum
\begin{equation}
    \mathbf{u}_i
    = \frac{\mathbf{p}_i}{m_s c}
    = \gamma_i\boldsymbol{\beta}_i,
    \qquad
    \gamma_i = \sqrt{1+\lVert\mathbf{u}_i\rVert^2},
\end{equation}
where $m_s$ and $q_s$ are the rest mass and charge of the species, $\mathbf{p}_i=\gamma_i m_s\mathbf{v}_i$ is the physical momentum, $\mathbf{v}_i$ is the particle velocity, $\boldsymbol{\beta}_i=\mathbf{v}_i/c$, and $c$ is the speed of light. Positions are expressed in meters, and the normalized momentum components are dimensionless, following the convention used in \OPAL{}~\cite{OPAL}.

The relativistic equations of motion are
\begin{align}
    \frac{\mathrm{d}\mathbf{r}_i}{\mathrm{d}t}
        &= c\frac{\mathbf{u}_i}{\gamma_i}, \\
    \frac{\mathrm{d}\mathbf{u}_i}{\mathrm{d}t}
        &= \frac{q_s}{m_s c}
           \left(
           \mathbf{E}_i
           + c\frac{\mathbf{u}_i}{\gamma_i}
             \times\mathbf{B}_i
           \right),
\end{align}
which correspond to the Lorentz-force equations written in terms of normalized mechanical momentum~\cite{Reiser:2008}. Here $t$ is the laboratory time, and $\mathbf{E}_i$ and $\mathbf{B}_i$ are the total laboratory-frame fields evaluated at the particle position $\mathbf{r}_i$ and time $t$. They contain prescribed accelerator fields and self-fields determined from the evolving particle distribution. Their calculation and combination are described in Secs.~\ref{sec:self-field} and \ref{sec:particle-field}.

In addition to the macroparticle ensemble, \OPALX{} advances a reference particle using the external accelerator fields. The reference particle does not contribute to the collective charge distribution; its trajectory defines the reference path through the lattice and supports the identification of active elements. Reference-path construction and element activation are described in Sec.~\ref{sec:architecture}, while element placement and external-field evaluation are discussed in Sec.~\ref{sec:components}.

\subsection{Time integration and operator splitting}\label{sec:time-integration}

\OPALX{} advances the particle ensemble in laboratory time using a drift--kick--drift scheme with a relativistic Boris momentum update~\cite{Zenitani:2018,OPALXManual}. For a particle present at the beginning of a step of duration $\Delta t=t^{n+1}-t^n$, the first half drift is
\begin{equation}
    \mathbf{r}_i^{n+1/2}
    = \mathbf{r}_i^n
      + \frac{c\Delta t}{2}
        \frac{\mathbf{u}_i^n}{\gamma_i^n},
    \qquad
    \gamma_i^n=\sqrt{1+\lVert\mathbf{u}_i^n\rVert^2}.
    \label{eq:integration-first-drift}
\end{equation}
The momentum remains unchanged during this drift. In the default \texttt{MIDPOINT} ordering, the self-fields are then recalculated from the half-step particle positions. External fields are evaluated at the half-step particle positions. RF-cavity fields use midpoint time, whereas time-dependent \texttt{MultipoleT} scaling currently uses the step-start time. The total fields supplied to the momentum update are denoted by $\mathbf{E}_i^*$ and $\mathbf{B}_i^*$; their calculation and transformation are described in Secs.~\ref{sec:self-field} and \ref{sec:particle-field}.

The Boris update splits the momentum advance into two half electric impulses separated by a magnetic rotation. The first electric impulse gives
\begin{equation}
    \mathbf{u}_i^-
    = \mathbf{u}_i^n
      + \frac{q_s\Delta t}{2m_s c}\mathbf{E}_i^*.
\end{equation}
The rotation parameters are
\begin{align}
    \gamma_i^- &= \sqrt{1+\lVert\mathbf{u}_i^-\rVert^2}, \\
    \boldsymbol{\tau}_i
        &= \frac{q_s\Delta t}{2m_s\gamma_i^-}\mathbf{B}_i^*,
    \qquad
    \boldsymbol{\eta}_i
        = \frac{2\boldsymbol{\tau}_i}
               {1+\lVert\boldsymbol{\tau}_i\rVert^2}.
\end{align}
Using an intermediate vector $\widetilde{\mathbf{u}}_i$, the rotation and second electric impulse are
\begin{align}
    \widetilde{\mathbf{u}}_i
        &= \mathbf{u}_i^-
           + \mathbf{u}_i^-\times\boldsymbol{\tau}_i, \\
    \mathbf{u}_i^+
        &= \mathbf{u}_i^-
           + \widetilde{\mathbf{u}}_i\times\boldsymbol{\eta}_i, \\
    \mathbf{u}_i^{n+1}
        &= \mathbf{u}_i^+
           + \frac{q_s\Delta t}{2m_s c}\mathbf{E}_i^*.
    \label{eq:integration-boris-kick}
\end{align}
The magnetic substep preserves the momentum magnitude, $\lVert\mathbf{u}_i^+\rVert=\lVert\mathbf{u}_i^-\rVert$, up to roundoff. The step is completed by the second half drift,
\begin{equation}
    \mathbf{r}_i^{n+1}
    = \mathbf{r}_i^{n+1/2}
      + \frac{c\Delta t}{2}
        \frac{\mathbf{u}_i^{n+1}}{\gamma_i^{n+1}},
    \label{eq:integration-second-drift}
\end{equation}
where $\gamma_i^{n+1}=\sqrt{1+\lVert\mathbf{u}_i^{n+1}\rVert^2}$.

The \texttt{PRESTEP} option changes the sampling of the collective fields: the self-field solve precedes the first half drift and uses the particle positions $\mathbf{r}_i^n$. The gathered fields are carried with the particles to the momentum update and combined with external fields evaluated at the half-step particle positions, using the element-dependent time sampling described above. This option retains the historical \OPAL{} space charge sampling point; the Boris update and position drifts are unchanged~\cite{OPALXManual}.

Time-dependent emission is inserted after the self-field solve. Newly created particles use the fraction of the step following their birth time, as described by Eq.~\eqref{eq:emission-dt}; they first contribute to the subsequent collective-field update. The placement of spin tracking and additional processes within the complete step is described in Sec.~\ref{sec:architecture}.

For sufficiently smooth prescribed fields evaluated consistently at the midpoint, the centered drift--Boris-kick--drift scheme is second-order accurate. The current step-start sampling of time-dependent \texttt{MultipoleT} scaling does not satisfy this temporal-centering condition, so second-order accuracy should not be assumed for such configurations. The temporal accuracy of a self-consistent calculation also depends on the sampling of the source distribution and the source-frame quantities used to construct the self-fields. Midpoint particle positions alone do not ensure that source velocities and Lorentz factors are centered in time. The accuracy of the coupled calculation must therefore be established through time-step refinement. The configured time-step schedule must resolve the relevant field variations, particle motion, emission, and collective dynamics.

\subsection{Spin dynamics and tracking}
The spin state of a spin-1/2 particle is described by a density matrix $\bm{\rho}$: a $2\times 2$ Hermitian, positive semidefinite matrix with unit trace \cite{Fano:1957}. It can always be written as
\begin{equation}
    \bm{\rho} = \sum_i p_i\, \bm{\chi}_i \bm{\chi}_i^{\dagger},
    \qquad p_i \ge 0, \quad \sum_i p_i = 1,
\end{equation}
where each
\begin{equation}
    \bm{\chi}_i =
    \begin{pmatrix}
        \alpha_i \\
        \beta_i
    \end{pmatrix}
    \in \mathbb{C}^2,
    \qquad |\alpha_i|^2 + |\beta_i|^2 = 1,
\end{equation}
is a normalized pure spin state with statistical weight $p_i$. All observable quantities depend only on $\bm{\rho}$.

The identity matrix $\bm{I}$ and the three Pauli matrices $\bm{\sigma}_x$, $\bm{\sigma}_y$, and $\bm{\sigma}_z$ form a basis of the real vector space of $2\times 2$ Hermitian matrices \cite{Fano:1957, Blum:2012}. Introducing the dimensionless spin-polarization vector $\bm{P}$ and writing $\bm{\sigma}=(\bm{\sigma}_x,\bm{\sigma}_y,\bm{\sigma}_z)$, the density matrix can therefore be written as
\begin{equation}
    \bm{\rho} = \frac{1}{2}\left(\bm{I} + \bm{P} \cdot \bm{\sigma}\right),
    \qquad \bm{P} \in \mathbb{R}^3, \quad |\bm{P}| \le 1,
\end{equation}
where the coefficient of $\bm{I}$ is fixed by the condition $\mathrm{Tr}(\bm{\rho}) = 1$,
and the bound $|\bm{P}| \le 1$ follows from the positivity of $\bm{\rho}$.

For $i \in \{x, y, z\}$, the expectation value of the Pauli operator $\bm{\sigma}_i$ is
\begin{equation}
    \langle \bm{\sigma}_i \rangle = \mathrm{Tr}(\bm{\rho}\bm{\sigma}_i) = P_i\,.
\end{equation}
The physical spin operator is $\bm{S}=(\hbar/2)\bm{\sigma}$, so that $\langle\bm{S}\rangle=(\hbar/2)\bm{P}$. More generally, for a unit vector $\bm{n}$, the observable $\bm{n}\cdot\bm{\sigma}$ yields the outcome $\pm 1$ with probability $\frac{1}{2}(1 \pm \bm{n}\cdot\bm{P})$, so that $\langle \bm{n}\cdot\bm{\sigma} \rangle = \bm{n}\cdot\bm{P}$. Since $\bm{\rho}$ is fully determined by $\bm{P}$, it suffices to track the evolution of $\bm{P}$ to statistically simulate the outcomes of spin measurements.

The evolution of the spin polarization is given by
\begin{equation}
    \frac{\mathrm{d}\langle \bm{\sigma}_j \rangle}{\mathrm{d}t}
    = \frac{i}{\hbar} \langle [\bm{H}, \bm{\sigma}_j] \rangle\,.
    \label{eq:ehrenfest}
\end{equation}
Since the Hamiltonian $\bm{H}$ is also a $2\times 2$ Hermitian matrix, it admits the decomposition
\begin{equation}
    \bm{H} = h_0 \bm{I} + \bm{h} \cdot \bm{\sigma}\,.
\end{equation}

The term $h_0 \bm{I}$ commutes with every matrix and therefore drops out of the commutator in Eq.~\eqref{eq:ehrenfest} and has no effect on the spin dynamics. We write the remaining vector as $\bm{h} = \frac{\hbar}{2}\bm{\Omega}$, where the factor $\hbar/2$ is chosen so that $\bm{\Omega}$ is the angular velocity of the precession derived below:
\begin{equation}
    \bm{H} = h_0 \bm{I} + \frac{\hbar}{2}\, \bm{\Omega} \cdot \bm{\sigma}\,.
\end{equation}
The Pauli matrices satisfy the commutation relation
\begin{equation}
    [\bm{\sigma}_k, \bm{\sigma}_l] = 2i\, \epsilon_{klm}\, \bm{\sigma}_m\,,
\end{equation}
where $\epsilon_{klm}$ is the Levi-Civita symbol and summation over repeated
indices is implied. The commutator of the Hamiltonian with $\bm{\sigma}_j$ is
therefore
\begin{equation}
    [\bm{H}, \bm{\sigma}_j]
    = \frac{\hbar}{2}\, \Omega_k\, [\bm{\sigma}_k, \bm{\sigma}_j]
    = i\hbar\, \epsilon_{kjl}\, \Omega_k\, \bm{\sigma}_l\,.
\end{equation}
Substituting this into Eq.~\eqref{eq:ehrenfest}, using $P_j = \langle \bm{\sigma}_j \rangle$, and treating $\bm{\Omega}$ as a classical quantity so that $\langle \Omega_k \bm{\sigma}_l \rangle = \Omega_k \langle \bm{\sigma}_l \rangle$, gives
\begin{equation}
    \frac{\mathrm{d}P_j}{\mathrm{d}t}
    = \frac{i}{\hbar}\, i\hbar\, \epsilon_{kjl}\, \Omega_k P_l
    = -\epsilon_{kjl}\, \Omega_k P_l
    = \epsilon_{jkl}\, \Omega_k P_l\,,
\end{equation}
where the last step uses the antisymmetry $\epsilon_{kjl} = -\epsilon_{jkl}$. The right-hand side is the $j$-th component of $\bm{\Omega} \times \bm{P}$, so
\begin{equation}
    \frac{\mathrm{d}\bm{P}}{\mathrm{d}t} = \bm{\Omega} \times \bm{P}\,.
\end{equation}
This describes a rotation of $\bm{P}$ about $\bm{\Omega}$ with angular frequency $|\bm{\Omega}|$, where $|\bm{P}|$ is conserved.

The evolution of the spin polarization of a particle with a classical orbit and no electric dipole moment is given by the Thomas--BMT equation \cite{Bargmann:1959, Thomas:1926, Jackson:1999}
\begin{align}
    \frac{\mathrm{d}\bm{P}}{\mathrm{d}t}
        &= \bm{\Omega}_{\mathrm{BMT}} \times \bm{P}\,, \label{eq:tbmt} \\
    \bm{\Omega}_{\mathrm{BMT}}
        &= -\frac{q}{m} \biggl[ \left(G + \frac{1}{\gamma}\right) \bm{B}
           \nonumber \\
        &\qquad - \frac{G\gamma}{\gamma+1} (\bm{\beta}\cdot\bm{B})\,\bm{\beta}
           \nonumber \\
        &\qquad - \left(G + \frac{1}{\gamma+1}\right)
           \frac{\bm{\beta}\times\bm{E}}{c} \biggr]\,, \label{eq:omega_bmt}
\end{align}
where $q$ and $m$ are the charge and mass of the particle, $G = (g-2)/2$ is its magnetic anomaly, $\gamma$ is the Lorentz factor, $\bm{\beta} = \bm{v}/c$, and $\bm{E}$ and $\bm{B}$ are the lab-frame fields at the position of the particle. The polarization $\bm{P}$ is defined in the particle rest frame, and $t$ is the lab time.
\vfill

\subsubsection{Rodrigues' Rotation}
Conventional explicit Runge--Kutta methods do not, in general, preserve the polarization magnitude exactly when integrating Eq.~\eqref{eq:tbmt}~\cite[Chap.~IV]{Hairer:2006}. For a pure rotation, the commonly used Boris update (the Boris-B variant) preserves the vector norm but introduces a finite-step phase error~\cite{Zenitani:2018}.

Rodrigues' rotation formula \cite{Rodrigues:1840, Dai:2015} gives the exact rotation of $\bm{P}$ about a unit vector $\bm{n}$ through an angle $\theta$.

To derive it, decompose $\bm{P}$ into components parallel and perpendicular to the rotation axis,
\begin{equation}
    \bm{P}_\parallel = (\bm{n}\cdot\bm{P})\,\bm{n}\,, \qquad
    \bm{P}_\perp = \bm{P} - (\bm{n}\cdot\bm{P})\,\bm{n}\,.
\end{equation}
The parallel component is unchanged by the rotation. The perpendicular component rotates in the plane orthogonal to $\bm{n}$, which is spanned by the orthogonal vectors $\bm{P}_\perp$ and $\bm{n}\times\bm{P}_\perp = \bm{n}\times\bm{P}$. The rotated vector $\bm{P}'$ is given by
\begin{align}
    \bm{P}' &= \bm{P}_\parallel + \bm{P}_\perp\cos\theta
              + (\bm{n}\times\bm{P})\sin\theta \nonumber \\
            &= \bm{P}\cos\theta + (\bm{n}\times\bm{P})\sin\theta
               \nonumber \\
            &\qquad + (\bm{n}\cdot\bm{P})\,\bm{n}\,(1-\cos\theta)\,.
               \label{eq:rodrigues}
\end{align}
Because the three terms $\bm{P}_\parallel$, $\bm{P}_\perp$ and $\bm{n}\times\bm{P}$ are mutually orthogonal,
\begin{equation}
    |\bm{P}'|^2 = |\bm{P}_\parallel|^2
                  + |\bm{P}_\perp|^2\left(\cos^2\theta + \sin^2\theta\right)
                = |\bm{P}|^2\,,
\end{equation}
so the norm is conserved for any $\theta$. For a nonzero precession vector $\bm{\Omega}_{\mathrm{BMT}}$ that is constant over a time step $\Delta t$, Eq.~\eqref{eq:rodrigues} with
\begin{equation}
    \bm{n} = \frac{\bm{\Omega}_{\mathrm{BMT}}}{|\bm{\Omega}_{\mathrm{BMT}}|}\,,
    \qquad
    \theta = |\bm{\Omega}_{\mathrm{BMT}}|\,\Delta t
\end{equation}
is the exact solution of Eq.~\eqref{eq:tbmt} over that step. When $|\bm{\Omega}_{\mathrm{BMT}}|=0$, the polarization is unchanged, $\bm{P}'=\bm{P}$.

\subsection{Self-field formulation}
\label{sec:self-field}
The space charge model uses an electrostatic approximation in a frame moving with the beam, following the formulation used in \OPAL{}~\cite{OPAL}. A relativistic beam can have a substantial magnetic self-field in the laboratory while its internal motion is nonrelativistic in this co-moving frame. Primes denote co-moving quantities, and $\mathbf{v}_b=c\boldsymbol{\beta}_b$ and $\gamma_b=(1-|\boldsymbol{\beta}_b|^2)^{-1/2}$ denote the frame velocity relative to the laboratory and its associated Lorentz factor, respectively. The self-fields obey Maxwell's equations,
\begin{align}
    \nabla'\cdot\mathbf{E}'_{\mathrm{self}} &= \frac{\rho'}{\epsilon_0},
    &\nabla'\cdot\mathbf{B}'_{\mathrm{self}} &= 0, \\
    \nabla'\times\mathbf{E}'_{\mathrm{self}}
        &= -\frac{\partial\mathbf{B}'_{\mathrm{self}}}{\partial t'},
    &\nabla'\times\mathbf{B}'_{\mathrm{self}}
        &= \mu_0\mathbf{J}'
        +\frac{1}{c^2}\frac{\partial\mathbf{E}'_{\mathrm{self}}}{\partial t'},
\end{align}
where $\rho'$ and $\mathbf{J}'$ are the charge and current densities, $\epsilon_0$ and $\mu_0$ are the vacuum permittivity and permeability, and $\mu_0\epsilon_0=c^{-2}$. If internal currents are sufficiently weak and the fields vary on timescales long compared with the light-crossing time of the bunch, magnetic and retardation effects can be neglected in this frame. Thus $\mathbf{B}'_{\mathrm{self}}\simeq0$ and $\nabla'\times\mathbf{E}'_{\mathrm{self}}\simeq0$, giving
\begin{equation}
    \mathbf{E}'_{\mathrm{self}}=-\nabla'\phi',
    \qquad \nabla'^2\phi'=-\frac{\rho'}{\epsilon_0}.
    \label{eq:self-poisson}
\end{equation}
Here $\rho'$ is measured in $\mathrm{C\,m^{-3}}$, $\phi'$ in volts, and $\mathbf{E}'_{\mathrm{self}}$ in $\mathrm{V\,m^{-1}}$. Boundary conditions are selected according to the solver and geometry; the free-space case is discussed in Sec.~\ref{sec:open-solver}.

For a boost along the local beam direction $z$, the Lorentz transformation of space-time coordinates is
\begin{equation}
    \begin{aligned}
        z'&=\gamma_b(z-v_b t), & x'&=x, \\
        t'&=\gamma_b\left(t-\frac{v_b z}{c^2}\right), & y'&=y.
    \end{aligned}
\end{equation}
The charge density transforms as $\rho'=\gamma_b(\rho-\mathbf{v}_b\cdot\mathbf{J}/c^2)$. For a distribution convected with the mean beam velocity, $\mathbf{J}\simeq\rho\mathbf{v}_b$, this reduces to $\rho'\simeq\rho/\gamma_b$. In the quasi-static approximation, a frozen laboratory snapshot is represented by stretching longitudinal separations by $\gamma_b$, with this density rescaling preserving total charge. This treatment neglects evolution between the different co-moving times associated with simultaneous laboratory particle positions.

The inverse Lorentz transformation, with $\mathbf{B}'_{\mathrm{self}}\simeq0$, gives~\cite{Jackson:1999}
\begin{align}
    \mathbf{E}_{\mathrm{self},\parallel}
        &=\mathbf{E}'_{\mathrm{self},\parallel},
    &\mathbf{E}_{\mathrm{self},\perp}
        &=\gamma_b\mathbf{E}'_{\mathrm{self},\perp}, \\
    \mathbf{B}_{\mathrm{self}}
        &=\frac{1}{c^2}\mathbf{v}_b\times\mathbf{E}_{\mathrm{self}},
    \label{eq:self-lorentz}
\end{align}
where parallel and perpendicular refer to $\mathbf{v}_b$, and $\mathbf{B}_{\mathrm{self}}$ is measured in tesla. Thus the laboratory magnetic self-field is retained even though it is neglected in the co-moving electrostatic solve.

If the longitudinal energy spread of a beam is large enough that the nonrelativistic approximation in the co-moving frame no longer holds, \OPALX{} can group particles into energy bins with their own mean velocity and Lorentz factor. Each bin contributes a quasi-static Poisson solve with longitudinal mesh stretching and charge-density normalization. Its fields are transformed back and summed. The approximation requires small internal velocities within each source group and slowly varying sources. Within each bin, the electrostatic approximation of Eq.~\eqref{eq:self-poisson} is applied, with internal currents neglected. \OPALX{} selects the histogram used to assign particles to energy bins using a cost function that balances field accuracy against computational cost and modeling error~\cite{Liemen:2025}. The co-moving frame is treated as instantaneously inertial; acceleration and curvature must be small enough on the bunch scale for this local self-field approximation to apply.

\subsection{Particle-field coupling}
\label{sec:particle-field}
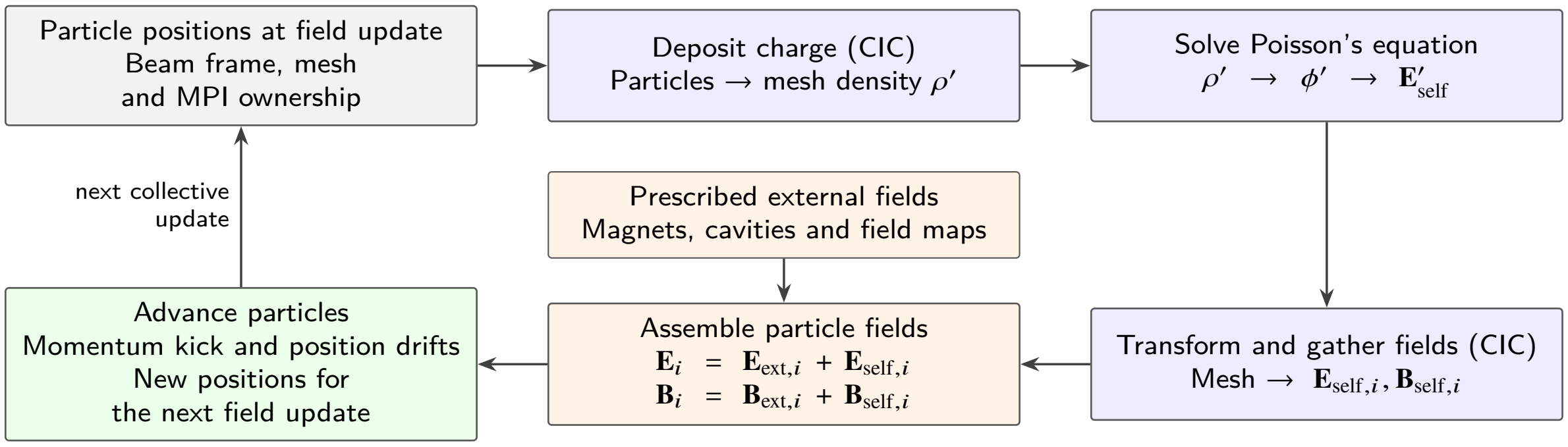
\begin{figure*}[t]
\centering
\begin{tikzpicture}[
    font=\sffamily\footnotesize,
    box/.style={draw=black!65, rounded corners=1pt, line width=0.5pt,
        align=center, text width=4.45cm, inner sep=4pt, minimum height=1.12cm},
    flow/.style={-{Stealth[length=1.9mm]},line width=0.65pt,draw=black!75}]
\node[box,fill=black!5] (particles) at (0,0)
    {Particle positions at field update\\Beam frame, mesh and MPI ownership};
\node[box,fill=blue!7] (deposit) at (5.45,0)
    {Deposit charge (CIC)\\Particles $\rightarrow$ mesh density $\rho'$};
\node[box,fill=blue!7] (solve) at (10.9,0)
    {Solve Poisson's equation\\$\rho'\rightarrow\phi'\rightarrow\mathbf{E}'_{\mathrm{self}}$};
\node[box,fill=blue!7] (gather) at (10.9,-3.0)
    {Transform and gather fields (CIC)\\Mesh $\rightarrow\mathbf{E}_{\mathrm{self},i},\mathbf{B}_{\mathrm{self},i}$};
\node[box,fill=orange!10] (total) at (5.45,-3.0)
    {Assemble particle fields\\$\mathbf{E}_i=\mathbf{E}_{\mathrm{ext},i}+\mathbf{E}_{\mathrm{self},i}$\\$\mathbf{B}_i=\mathbf{B}_{\mathrm{ext},i}+\mathbf{B}_{\mathrm{self},i}$};
\node[box,fill=green!8] (push) at (0,-3.0)
    {Advance particles\\Momentum kick and position drifts\\New positions for the next field update};
\node[draw=black!65,rounded corners=1pt,fill=orange!10,
    align=center,text width=4.45cm,inner sep=4pt] (external) at (5.45,-1.5)
    {Prescribed external fields\\Magnets, cavities and field maps};
\draw[flow] (particles.east) -- (deposit.west);
\draw[flow] (deposit.east) -- (solve.west);
\draw[flow] (solve.south) -- (gather.north);
\draw[flow] (gather.west) -- (total.east);
\draw[flow] (external.south) -- (total.north);
\draw[flow] (total.west) -- (push.east);
\draw[flow] (push.north) -- (particles.south)
    node[midway,left,align=right,font=\sffamily\scriptsize]
        {next collective\\update};
\end{tikzpicture}
\caption{Electrostatic particle-in-cell loop in \OPALX{}.
Charge deposition, the Poisson solve, and field gathering connect
particles to the computational mesh; particle advancement closes the loop.
For relativistic source bins, each bin is solved in its co-moving frame,
and its electric and magnetic fields are transformed and accumulated before
gathering. Prescribed external fields enter after the self-field calculation. The diagram shows the numerical data flow; the actual half-drift,
emission, and optional-process ordering is shown in
Fig.~\ref{fig:program-workflow}.}
\label{fig:pic-loop}
\end{figure*}

The PIC method closes the self-consistent coupling between beam motion and collective fields through the loop illustrated in Fig.~\ref{fig:pic-loop}. At each collective update, the current particle positions define the charge deposited on the mesh. The Poisson solver selected by the user computes the mesh field, which is gathered to the particles and combined with the prescribed accelerator fields. Particle advancement changes the distribution that supplies the next field calculation. The field solve is therefore one stage of a repeated particle--mesh--particle cycle, rather than a calculation with a fixed source~\cite{IPPL,OPAL}.
%

In the continuous formulation, the total laboratory fields at each particle position are
\begin{align}
    \mathbf{E}_i
        &=\mathbf{E}_{\mathrm{ext}}(\mathbf{r}_i,t)
          +\mathbf{E}_{\mathrm{self}}(\mathbf{r}_i,t), \\
    \mathbf{B}_i
        &=\mathbf{B}_{\mathrm{ext}}(\mathbf{r}_i,t)
          +\mathbf{B}_{\mathrm{self}}(\mathbf{r}_i,t).
\end{align}
The sampling positions and times used in the discrete integration are specified in Sec.~\ref{sec:time-integration}. The external fields are supplied by magnets, RF cavities, and other accelerator components, using analytical models or prescribed field maps. Particle coordinates are transformed to each active element's local frame for field evaluation, and the resulting vectors are rotated back to a common frame and added. Overlapping elements therefore contribute to the same external-field sum. Unlike the self-consistent fields, these prescribed fields do not depend on the deposited beam charge.

For the three-dimensional Cartesian PIC calculation, the self-fields are obtained on a computational mesh and interpolated to the particles through \IPPL{}~\cite{IPPL}. The computational domain is a rectangular box in beam-aligned coordinates. In the bunch-fitted configuration, its bounds follow the global particle-position extrema, enlarged by a configurable margin; the chosen mesh dimensions determine the uniform spacing in each direction. A fixed box can instead be prescribed for the open-boundary solver. This spatial translation and rotation into beam-aligned coordinates is distinct from the Lorentz transformation used for a source bin.

Charge deposition uses cloud-in-cell~(CIC) weights: a particle contributes to the eight neighboring mesh points, with weights given by products of linear weights in each coordinate. The deposited charge is normalized by the cell volume and, where applicable, by the Lorentz factor of the source bin. The selected solver then computes the co-moving electric field described in Sec.~\ref{sec:self-field}. For relativistic source bins, the laboratory electric and magnetic fields are formed and accumulated on the mesh before gathering to the particles.

Writing $W_{ia}$ for the CIC weight linking particle $i$ to mesh point $a$, the gather operation is
\begin{equation}
    \mathbf{E}_{\mathrm{self},i}=\sum_a W_{ia}\mathbf{E}_{\mathrm{self},a},
    \qquad
    \mathbf{B}_{\mathrm{self},i}=\sum_a W_{ia}\mathbf{B}_{\mathrm{self},a}.
\end{equation}
The weights satisfy $\sum_a W_{ia}=1$ and use the same interpolation shape as charge deposition, giving trilinear field interpolation in three dimensions. Both self-fields are expressed in the same frame as the external fields before their sum is applied in the Lorentz-force update. The collective Lorentz force includes both $q_s\mathbf{E}_{\mathrm{self},i}$ and $q_s\mathbf{v}_i\times\mathbf{B}_{\mathrm{self},i}$; the integration scheme and ordering of kicks are discussed in Sec.~\ref{sec:time-integration}. The resulting particle state closes the PIC loop: before the next charge deposition, the beam-aligned coordinates, domain bounds, and particle ownership are updated as required. \IPPL{} handles mesh-data communication and particle migration between MPI subdomains so that deposition and gathering use a consistent distributed mesh. The density and self-fields are recomputed from the evolving source distribution at each collective update. For the default \texttt{MIDPOINT} ordering, these source positions follow the first half drift; emission takes place after this solve, so newly emitted particles contribute to the following collective update. Section~\ref{sec:architecture} gives the full tracking sequence, including external-field evaluation, optional processes, and diagnostics.

\subsection{Grid-based and hybrid solvers}
The space charge solvers in \OPALX{} are built on \IPPL{}, which provides common distributed field data structures and performance-portable numerical kernels~\cite{IPPL}. The following subsections describe the three-dimensional open-boundary and 2.5D particle--mesh Poisson solvers, followed by the P3M method, which combines a mesh-based long-range calculation with direct short-range particle--particle interactions. The supported boundary conditions are also discussed.

\subsubsection{Open-boundary Poisson solver}
\label{sec:open-solver}
The \texttt{OPEN} solver evaluates the electrostatic field of a spatially localized charge distribution in free space. It solves Eq.~\eqref{eq:self-poisson} with the boundary condition $\phi'(\mathbf{r}')\rightarrow0$ as $|\mathbf{r}'|\rightarrow\infty$. The solution in the co-moving frame is the convolution
\begin{equation}
    \phi'(\mathbf{r}')=\frac{1}{\epsilon_0}
    \int_{\mathbb{R}^3}
    G(\mathbf{r}'-\mathbf{s}')\rho'(\mathbf{s}')\,\mathrm{d}^3s',
\end{equation}
with $G(\mathbf{r}')=1/(4\pi|\mathbf{r}'|)$.

\OPALX{} uses the Hockney--Eastwood algorithm implemented in \IPPL{}~\cite{Hockney:1988, Mayani:2025}. On a uniform Cartesian mesh with $N_x\times N_y\times N_z$ points and spacings $h_x,h_y,h_z$, the charge density is zero-padded onto a $2N_x\times2N_y\times2N_z$ mesh. The Green function is extended by reflection to represent all signed source--observer separations. Fast Fourier transforms~(FFTs) then evaluate the resulting cyclic convolution, and the solution is restricted to the original mesh. This embedding reproduces the discrete free-space convolution without introducing periodic images into the physical domain. The convolution includes the cell-volume factor $h_xh_yh_z$.

Two Green-function discretizations are available. The \texttt{STANDARD} option samples $G$ at mesh separations and regularizes its singular value at the origin. The default \texttt{INTEGRATED} option instead uses the cell average of $G$, assuming uniform charge density within each source cell~\cite{Qiang:2006}. This gives a finite contribution from the source cell itself and is particularly useful for beams with large aspect ratios. The co-moving field $\mathbf{E}'_{\mathrm{self}}=-\nabla'\phi'$ is calculated by Fourier differentiation, followed by an inverse transform for each component.

The FFTs are provided through \IPPL{} by heFFTe~\cite{heFFTe}; MPI provides the communication required by the distributed transforms and field redistribution, while \Kokkos{} executes the local mesh kernels on CPUs or GPUs~\cite{Mayani:2025}. For $N_g=N_xN_yN_z$, the solve has $\mathcal{O}(N_g\log N_g)$ computational complexity, with additional storage for the enlarged mesh. Accuracy remains limited by mesh resolution, charge deposition, field interpolation, and particle statistics; using FFTs does not by itself imply spectral convergence.

\subsubsection{A 2.5D Poisson Solver (FFT2D5)}

The 2.5D solver calculates space charge fields through a sequence of two-dimensional transverse Poisson solves coupled to an approximate longitudinal-field model~\cite{OPALXManual}. The mesh follows a reference path expressed in Frenet--Serret coordinates, restricting the computational domain to the region around the beam. This can reduce mesh storage for extended or curved beam trajectories. For $N_s$ longitudinal slices with $N_x\times N_y$ transverse mesh points per slice, the transverse FFT work scales as $\mathcal{O}\!\left(N_sN_xN_y\log(N_xN_y)\right)$, compared with $\mathcal{O}\!\left(N_sN_xN_y\log(N_sN_xN_y)\right)$ for a three-dimensional FFT calculation with the same mesh-point counts. The method can also use a comparatively coarse longitudinal resolution when the longitudinal charge distribution varies slowly. It assumes that the beam is much longer than its transverse dimensions and is resolved by a sufficient number of longitudinal slices.

\begin{figure}[ht]
    \centering
    \includegraphics[width=\columnwidth]{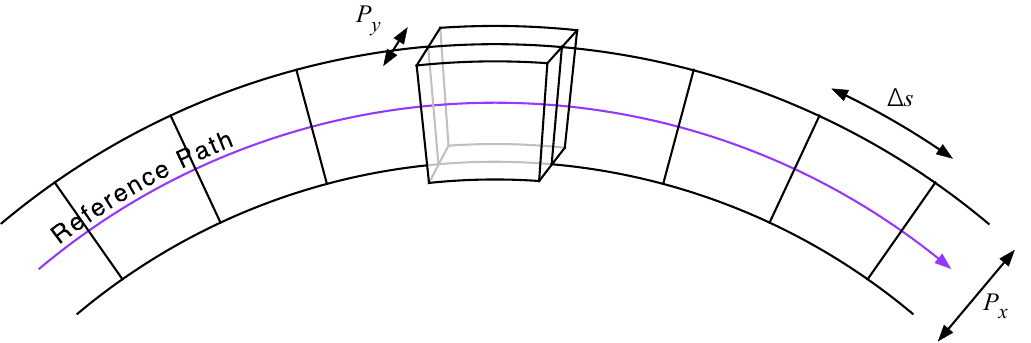}
    \caption{The Frenet--Serret reference path is divided into
    longitudinal slices, each discretized on a two-dimensional
    transverse grid. The mesh covers the region around the
    reference path.}
    \label{fig:CellDetermination}
\end{figure}

A reference path defines the Frenet--Serret coordinate system. In the current implementation, the path is assumed to lie in a horizontal plane. The coordinate $s$ measures distance along the path, $y$ is vertical, and $x$ is the horizontal transverse coordinate. The parameters $P_x$ and $P_y$ specify the transverse extent of the computational domain, as illustrated in Fig.~\ref{fig:CellDetermination}. The slice width is $\Delta s$, and each slice is discretized using transverse cell sizes $\Delta x$ and $\Delta y$. The reference path can be supplied by the \OPALX{} \orbitt{} or read from an external file.

\begin{figure}[ht]
    \centering
    \includegraphics[width=\columnwidth]{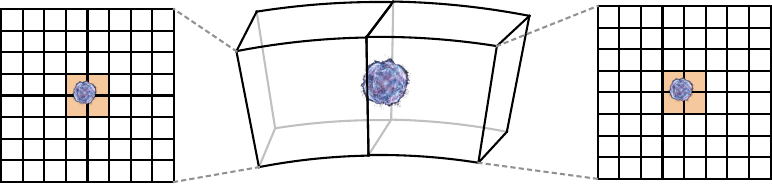}
    \caption{CIC deposition distributes the particle charge,
    represented by the blue cloud, among four transverse grid
    cells in each of two adjacent slices. The corresponding
    interpolation weights are used to gather fields to the
    particle position.}
    \label{fig:CiCAlgorithm}
\end{figure}

Particle positions and momenta are first transformed into Frenet--Serret coordinates and then boosted into the beam's co-moving frame. Charge deposition uses CIC weights, with a trilinear scheme distributing each particle's charge among four transverse grid cells in each of the two adjacent slices, as illustrated in Fig.~\ref{fig:CiCAlgorithm}. These operations are parallelized over particles and can execute on a GPU.

The transverse electric-field components $E_x$ and $E_y$ are calculated independently in each slice. The current implementation supports open transverse boundary conditions and uses a two-dimensional FFT Poisson solve with the Hockney--Eastwood doubled-domain construction~\cite{Hockney:1988}. The transverse fields are obtained from the potential gradients. The local mesh operations are parallelized over grid cells and can execute on a GPU.

The longitudinal electric field is estimated from the derivative of the charge line density with respect to $z$, the longitudinal coordinate relative to the bunch. This coordinate is distinct from the path coordinate $s$ along the machine. The line density is calculated from the charge density in each slice, and its longitudinal derivative is approximated using central differences. The field is given by
\begin{equation}
    E_z(z)=-\frac{g}{4\pi\varepsilon_0\gamma^2}
    \frac{\mathrm{d}\lambda(z)}{\mathrm{d}z},
\end{equation}
where $\lambda(z)$ is the charge line density and $g$ is a constant determined by the boundary conditions. This calculation is parallelized over slices, so the available parallelism depends on the number of longitudinal slices.

The fields at each particle position are obtained using CIC interpolation from the adjacent grid cells and slices. The inverse Lorentz transformation and the transformation from Frenet--Serret to Cartesian coordinates return the electric and magnetic fields to the laboratory frame for use in particle tracking. These operations are parallelized over particles and can execute on a GPU.

\subsubsection{Particle--Particle Particle--Mesh~(P3M)}
\label{sec:p3m}
CIC deposition and field interpolation smooth the charge distribution, limiting the accuracy of mesh-based forces at separations comparable to or smaller than mesh spacing $h$. Conceptually, the field can be separated into a smooth mean field and a fluctuating contribution:
\begin{equation}
    \mathbf{E}_i 
    = \underbrace{\expval{ \mathbf{E} } \qty(\mathbf{r}_i)}_{\text{mean field, smoothed }\rho} 
    + \underbrace{\delta \mathbf{E}_i }_\text{discreteness, correlations}.
\end{equation}
Conventional grid-based FFT solvers do not accurately resolve the short-range fluctuations, $\delta \mathbf{E}_i$, associated with discrete Coulomb interactions because of mesh smoothing and the macroparticle representation~\cite{Mitchell:2015}. These interactions give rise to IBS~\cite{Piwinski:1974, Bjorken:1983}, which can significantly affect beam quality in dense, cold, high-brightness beams. In storage rings, IBS can limit the equilibrium emittance and energy spread~\cite{Zampetakis:2024, Bane:2002}. In linacs and high-brightness photoinjectors, it can increase the energy spread beyond predictions from space charge models that omit collisional effects~\cite{Prat:2022, Lucas:2026, DiMitri:2020}. \OPALX{} provides a GPU-enabled P3M implementation to include direct short-range particle interactions alongside the mesh-based field calculation.

The method separates the Coulomb Green function into long-range and short-range contributions~\cite{Hockney:1988}:
\begin{equation}
    G(r)
    = \underbrace{\frac{\mathrm{erf}(\alpha r)}{4\pi r}}_{G_L(r)}
    + \underbrace{\frac{\mathrm{erfc}(\alpha r)}{4\pi r}}_{G_S(r)},
\end{equation}
where $r>0$ is the particle separation and $\alpha>0$ is an inverse-length splitting parameter. The long-range contribution reuses the open-boundary FFT solver described in Sec.~\ref{sec:open-solver}, replacing its Green function with the long-range kernel and the regularization used by the implementation. This retains the distributed FFT infrastructure supplied by \IPPL{}.

The total field is $\mathbf{E}=\mathbf{E}_{\mathrm{mesh}}+\mathbf{E}_{\mathrm{pp}}$, with the short-range contribution calculated directly as
\begin{equation*}
    \mathbf{E}_{\mathrm{pp},i}
    =\frac{1}{4\pi\epsilon_0}
        \sum_{\substack{j\neq i\\ r_{ij}<r_c}}
        q_j\,\hat{\mathbf{r}}_{ij}\,
        \frac{f(\alpha r_{ij})}{r_{ij}^{2}},
\end{equation*}
where $f(x)=\mathrm{erfc}(x)+\tfrac{2x}{\sqrt{\pi}}e^{-x^{2}}$, $\mathbf{r}_{ij}=\mathbf{r}_i-\mathbf{r}_j$,  
$r_{ij}=|\mathbf{r}_{ij}|$, and $\hat{\mathbf{r}}_{ij}=\mathbf{r}_{ij}/r_{ij}$. Here $q_j$ is the charge carried by source particle $j$, and $r_c$ is the short-range cutoff radius. For an approximately uniform numerical particle density $n=N/V$, each particle has about $\tfrac{4\pi}{3}r_c^3n$ neighbors. The total pair work therefore scales as $N^2r_c^3/V$, where $N$ is the number of numerical particles and $V$ is the volume they occupy. \OPALX{} provides the \texttt{LOADBALANCINGPAIRWEIGHT} option to account for pair-interaction work when balancing the load across MPI ranks through orthogonal recursive bisection~(ORB)~\cite{OPALXManual}. 

The pair summation can be parallelized once candidate neighbors have been identified. Efficient neighbor search and communication are therefore important parts of the implementation. Each MPI rank stores ghost copies of nearby particles needed for interactions within $r_c$. Particles are sorted into search cells with widths of at least $r_c$, allowing each particle to search its own cell and the 26 adjacent cells. Pair contributions are accumulated using atomic operations. The domain decomposition constrains $r_c$ to at most half the local subdomain extent in every direction. 

In addition to open boundaries, \OPALX{} supports periodic P3M calculations. In this case, the periodic mesh solver uses the long-range kernel, while periodic image ghosts provide short-range interactions across the domain boundaries.

The split is characterized by the dimensionless parameters $\alpha h_\mathrm{max}$ and $\alpha r_c$, which control mesh-resolution and cutoff-truncation effects, respectively. With $\alpha = 2/r_c$, the ratio $\kappa = r_c/h$ expresses the cutoff in units of the mesh spacing $h$. The cutoff can be specified as an absolute length through \texttt{RCUT} or relative to the mesh spacing through \texttt{RCUT\_CELLS}. In the latter case, it is updated as the bunch-fitted mesh evolves. The cutoff must also satisfy the local-subdomain restriction stated above.

An image-field correction for a source plane at a prescribed $z$ coordinate uses a shifted Green function for the long-range contribution. The boundary treatment and its applicability conditions are discussed in Sec.~\ref{sec:solvers:boundary-conditions}.

The physical interpretation of the short-range dynamics depends on the macroparticle weights. Direct pair forces between macroparticles representing many physical particles generally do not reproduce the physical beam's binary-collision dynamics and can alter the evolution of second-order moments. For electron-beam calculations intended to resolve individual electron collisions, each numerical particle must represent one electron. This distinction is essential when interpreting collisional heating and emittance or energy-spread growth.

\subsubsection{Boundary conditions supported by \OPALX{}}
\label{sec:solvers:boundary-conditions}
For the three-dimensional mesh and P3M solvers, the boundary conditions of the field problem must be distinguished from the handling of the numerical particle domain. By default, the mesh bounds are updated to follow the extent of the bunch. This adaptation does not determine the physical boundary conditions, which are selected through the field solver. The \texttt{TYPE=FFT} backend solves a periodic Poisson problem, whereas \texttt{TYPE=OPEN} uses the free-space solver described in Sec.~\ref{sec:open-solver}, with the potential tending to zero at infinity. For the periodic FFT solve, the mean charge density is subtracted, corresponding to a uniform neutralizing background~\cite{OPALXManual}.

In periodic P3M calculations, the mesh solve uses the long-range kernel $G_L$ defined in Sec.~\ref{sec:p3m}. Periodic copies of particles are included in the short-range sum through ghost particles exchanged across domain boundaries. For open P3M calculations, the long-range contribution is calculated with the open-boundary FFT solver using $G_L$, and no periodic particle copies are included. These periodic copies are distinct from the opposite-sign image charges used to represent a conducting boundary.

A grounded planar cathode is represented by an image-charge construction. In the frame in which the electrostatic problem is solved, let the plane be $z=z_p$. All quantities in the following expressions refer to this frame. For a real charge $q$ at $(x,y,z)$, an image charge $-q$ is placed at $(x,y,2z_p-z)$. The corresponding image potential is
\begin{equation}
    \phi_\mathrm{image}(x,y,z)
    =-\phi_\mathrm{real}(x,y,2z_p-z),
\end{equation}
so that $\phi_\mathrm{real}+\phi_\mathrm{image}=0$ on the plane. This construction describes an infinite grounded plane.

For the open-boundary mesh solver, the image contribution can be obtained either by depositing the mirrored, opposite-sign charge distribution explicitly or by using a shifted Green function~\cite{OPALXManual}. The explicit construction requires the mesh to contain both the bunch and its image. The shifted-Green method avoids resolving the empty region between them~\cite{Qiang:2006}. A convolution with a kernel shifted by a displacement $\mathbf{s}$ defines a potential,
\begin{equation}
    \phi_\mathrm{shifted}(\mathbf{r})
    =\frac{1}{\epsilon_0}\int_{\mathbb{R}^3}
    G(\mathbf{r}-\mathbf{s}-\boldsymbol{\xi})
    \rho(\boldsymbol{\xi})\,\mathrm{d}^3\xi,
\end{equation}
where $G(\mathbf{r})=1/(4\pi|\mathbf{r}|)$ and $\boldsymbol{\xi}$ is the source integration coordinate. The associated field is $\mathbf{E}_\mathrm{shifted}=-\nabla\phi_\mathrm{shifted}$. The shift is set by the relative positions of the mesh and the cathode plane. Reflection and the image-charge sign are then applied to obtain the image contribution, consistently with the potential relation above.

For open P3M calculations, the cathode correction uses the shifted long-range kernel $G_L$ and adds the resulting image field to the mesh and direct-pair fields of the real bunch. Short-range interactions with the image charges are omitted. Within the cutoff approximation, these omitted interactions vanish at the real-particle positions if all real-particle--image separations exceed $r_c$. A sufficient condition is that the entire bunch lie more than $r_c$ from the plane. The error associated with omitting short-range boundary contributions has been analyzed by~\textcite{wyssling_boundary_2021}. Near the cathode, particularly during emission, this separation condition can fail, and the error from omitting short-range image interactions must be assessed.

\section{Program structure and simulation workflow}
\label{sec:architecture}
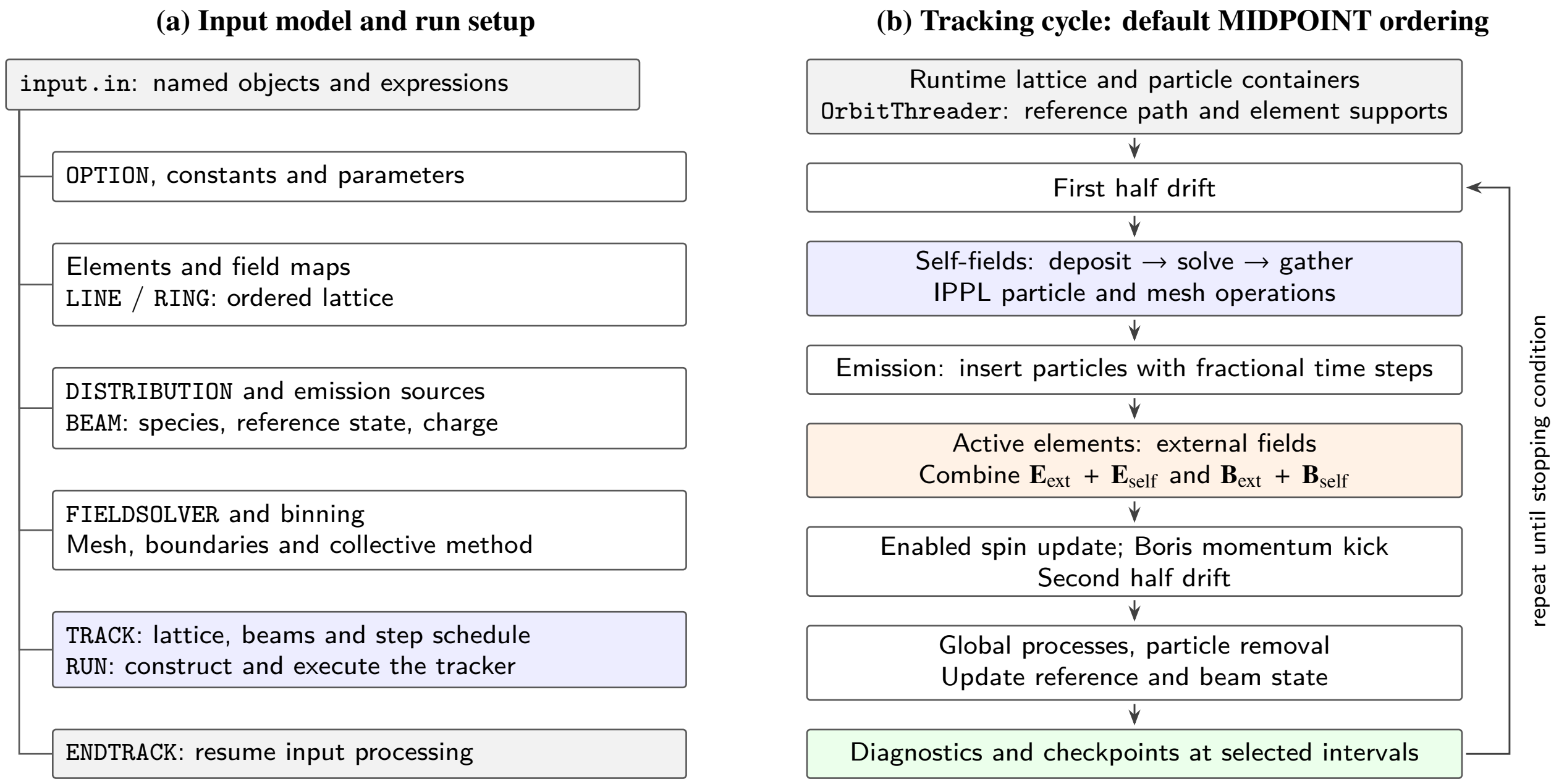
\begin{figure*}[t]
\centering
\begin{minipage}[t]{0.50\textwidth}
\centering
\textbf{(a) Input model and run setup}\par\vspace{2mm}
\begin{tikzpicture}[
    font=\sffamily\footnotesize,
    box/.style={draw=black!65, rounded corners=1pt, line width=0.5pt,
        align=left, text width=6.5cm, inner sep=4pt,
        anchor=north west},
    branch/.style={draw=black!65, line width=0.5pt}]
\def\inputgap{4.33mm}

\node[box,fill=black!5] (input) at (0,0)
    {\texttt{input.in}: named objects and expressions};
\node[box] (options) at ($(input.south west)+(5mm,-\inputgap)$)
    {\texttt{OPTION}, constants and parameters};
\node[box] (lattice) at ($(options.south west)+(0,-\inputgap)$)
    {Elements and field maps\\\texttt{LINE} / \texttt{RING}: ordered lattice};
\node[box] (beam) at ($(lattice.south west)+(0,-\inputgap)$)
    {\texttt{DISTRIBUTION} and emission sources\\\texttt{BEAM}: species, reference state, charge};
\node[box] (solver) at ($(beam.south west)+(0,-\inputgap)$)
    {\texttt{FIELDSOLVER} and binning\\Mesh, boundaries and collective method};
\node[box,fill=blue!7] (track) at ($(solver.south west)+(0,-\inputgap)$)
    {\texttt{TRACK}: lattice, beams and step schedule\\\texttt{RUN}: construct and execute the tracker};
\node[box,fill=black!5] (end) at ($(track.south west)+(0,-\inputgap)$)
    {\texttt{ENDTRACK}: resume input processing};

\foreach \target in {options,lattice,beam,solver,track,end}
    \draw[branch] (input.south west) ++(0.15,0) |- (\target.west);
\end{tikzpicture}
\end{minipage}\hfill
\begin{minipage}[t]{0.50\textwidth}
\centering
\textbf{(b) Tracking cycle: default MIDPOINT ordering}\par\vspace{2mm}
\begin{tikzpicture}[
    font=\sffamily\footnotesize,
    box/.style={draw=black!65, rounded corners=1pt, line width=0.5pt,
        align=center, text width=6.8cm, inner sep=3pt,
        minimum height=0.52cm, anchor=north},
    flow/.style={-{Stealth[length=1.7mm]}, line width=0.55pt,
        draw=black!75, shorten <=1pt, shorten >=1pt}]
\def\workflowgap{3mm}

\node[box,fill=black!5] (setup) at (0,0)
    {Runtime lattice and particle containers\\\texttt{OrbitThreader}: reference path and element supports};
\node[box] (drift) at ($(setup.south)+(0,-\workflowgap)$)
    {First half drift};
\node[box,fill=blue!7] (self) at ($(drift.south)+(0,-\workflowgap)$)
    {Self-fields: deposit $\rightarrow$ solve $\rightarrow$ gather\\\IPPL{} particle and mesh operations};
\node[box] (emission) at ($(self.south)+(0,-\workflowgap)$)
    {Emission: insert particles with fractional time steps};
\node[box,fill=orange!10] (external) at ($(emission.south)+(0,-\workflowgap)$)
    {Active elements: external fields\\Combine $\mathbf{E}_{\mathrm{ext}}+\mathbf{E}_{\mathrm{self}}$ and $\mathbf{B}_{\mathrm{ext}}+\mathbf{B}_{\mathrm{self}}$};
\node[box] (push) at ($(external.south)+(0,-\workflowgap)$)
    {Enabled spin update; Boris momentum kick\\Second half drift};
\node[box] (process) at ($(push.south)+(0,-\workflowgap)$)
    {Global processes, particle removal\\Update reference and beam state};
\node[box,fill=green!8] (output) at ($(process.south)+(0,-\workflowgap)$)
    {Diagnostics and checkpoints at selected intervals};

\foreach \from/\to in {setup/drift,drift/self,self/emission,emission/external,external/push,push/process,process/output}
    \draw[flow] (\from.south) -- (\to.north);

\coordinate (return-x) at ($(setup.east)+(5mm,0)$);
\coordinate (return-bottom) at (output.east -| return-x);
\coordinate (return-top) at (drift.east -| return-x);
\draw[flow] (output.east) -- (return-bottom)
    -- (return-top) -- (drift.east);
\node[font=\sffamily\scriptsize, rotate=90, anchor=north]
    at ($(return-bottom)!0.5!(return-top)+(1mm,0)$)
    {repeat until stopping condition};
\end{tikzpicture}
\end{minipage}
\caption{Input organization and tracking workflow of \OPALX{}.
(a) Named input objects define the model and bind it to a tracking run;
the branches group configuration responsibilities rather than prescribing
a universal declaration order. (b) The default collective update uses
positions after the first half drift. Emission follows this solve, so new
particles enter the next collective update. The \texttt{PRESTEP} alternative
moves the self-field calculation before the first half drift.
Optional processes and diagnostics are shown in their place in the cycle.}
\label{fig:program-workflow}
\end{figure*}

Figure~\ref{fig:program-workflow} summarizes the input organization and tracking cycle. The following subsections describe how these stages connect the accelerator model, particle ensemble, field calculations, and output.
%

\subsection{Input and lattice construction}
A simulation is specified through a MAD-inspired, statement-oriented input language retained from \OPAL{}. Named objects describe the beam, particle distributions and emission sources, accelerator elements, lattice, and collective-field configuration. Expressions and nested \texttt{LINE} definitions allow parameters and lattice segments to be reused. A \texttt{RING} defines an ordered ring sequence; its declaration alone does not place the elements or enforce geometric closure. Element poses and field maps determine the spatial accelerator model, as discussed in Sec.~\ref{sec:components}. The available commands and their current compatibility with \OPAL{} are documented in the \OPALX{} manual~\cite{OPALXManual}.

The parser resolves these objects through a common registry. A \texttt{TRACK} block selects the lattice and beam definitions and supplies the time-step, step-count, and stopping-position schedules. Within this block, \texttt{RUN} selects the parallel tracker and field solver; \texttt{ENDTRACK} returns to ordinary input processing. Setup constructs the runtime beamline, particle containers, sources, and a validated space charge configuration. Each selected beam has its own particle container and reference state. The solver and bunch persist for the run, while the tracker coordinates their use at each step.

\subsection{Tracking workflow}
Before ensemble tracking, the \orbitt{} advances a reference particle through the external fields and records the encountered element supports in an index map. Queries along this reference path provide candidate active elements during tracking. Element-local field bounds then determine which contributions apply to each particle; overlapping supports can contribute simultaneously. Supported analytic ring configurations also use spatial field selection. The reference orbit therefore organizes lattice access without replacing particle-level field evaluation or including the collective force.

Normal parallel bunch tracking uses a Boris drift--kick--drift cycle. In the default \texttt{MIDPOINT} ordering, particles first undergo a half drift, after which their self-fields are recalculated. Emission sources are then evaluated, and external fields are evaluated at the half-step particle positions, with the element-dependent time sampling described in Sec.~\ref{sec:time-integration}. Newly emitted particles receive an individual time increment for the fraction of the step during which they exist. Enabled spin tracking uses the assembled fields, and the momentum kick and second half drift complete the orbital step. Configured global processes, such as decay and daughter creation, are applied subsequently; invalid and lost particles are removed, reference and beam state are updated, and diagnostics are written at their requested intervals.

The \texttt{PRESTEP} option instead evaluates space charge before the first half drift and carries the gathered fields with the particles to the kick, reproducing the historical \OPAL{} sampling point. The configured step schedule and stopping conditions determine progression through the run. Numerical properties of the integration and collective models belong to Sec.~\ref{sec:physics}; their ordering is explicit here because it affects the interpretation and reproduction of a calculation.

\subsection{Coupling of external and collective fields}
The tracker combines two field-evaluation paths. Active accelerator elements evaluate prescribed fields in their local coordinates and add the transformed contributions to the particle field arrays. The space charge module obtains the collective contribution from the current particle distribution and returns it in the tracker coordinates. Both paths feed the same particle push, using the electric and magnetic field sums defined in Sec.~\ref{sec:particle-field}.

At each collective update, the tracker supplies particle-container activity, time-step information, and coordinate transformations through a common solve context. The selected space charge algorithm manages its domain, deposition, field solution, and gathering. Poisson adapters connect its field storage to the corresponding \IPPL{} backend. This separation allows the collective method to be configured independently of element-field evaluation, with \texttt{TYPE=NONE} selecting tracking without collective fields. The supported source containers and boundary conditions are properties of the selected algorithm. The formulation, interpolation, and solver equations remain in Sec.~\ref{sec:physics}.

\subsection{Diagnostics and data analysis}
Diagnostics combine beam-wide observables with localized measurements. Statistics record quantities such as particle number, energy, beam centroids, rms sizes, and emittances. Monitors collect bunch crossings at selected lattice locations, while probes record crossings through specified spans. Aperture and loss handling remove particles from the tracked ensemble and provide loss records where supported by the corresponding element. Output frequencies are configurable independently for statistics, phase-space snapshots, and checkpoints.

Particle phase space is written in Hierarchical Data Format version~5~(HDF5), and beam statistics are provided in Self-Describing Data Sets~(SDDS) files. Optional particle-field and rank output, together with external-field sampling through \texttt{DUMPEMFIELDS}, supports inspection of field evaluation and parallel particle ownership. Beam moments and selected diagnostics are evaluated in situ during tracking, reducing the need to reconstruct them from full particle dumps. Exported particle and field data, together with geometry output in Visualization Toolkit~(VTK) format, support visualization and subsequent analysis.

HDF5 checkpoints preserve particle and optional spin data, reference states and coordinate transforms, RF phases, and the current time and step-schedule position. Restart redistributes the saved particles over the participating MPI ranks and resumes compatible output streams. Checkpointing is independent of diagnostic phase-space dumping. Current restart support excludes time-dependent emission and stochastic decay, whose sampler or random-pool states are not saved. Changes of MPI rank count or rebuilt iterative-solver state can also change the final floating-point bits of a continuation~\cite{OPALXManual}.

\subsection{Extensibility and reproducibility}
New accelerator components extend the common element representation, including field support, coordinate handling, and field evaluation. Parser-visible objects are registered with their attributes so that they can be constructed from input statements. New collective methods implement the space charge algorithm interface; Poisson backends are connected through adapters to \IPPL{}. These interfaces separate model implementation from the overall tracking cycle. Extensions require appropriate unit and regression tests, documentation of assumptions and units, and verification on the intended execution backends.

\OPALX{} is distributed as open-source software under the GNU General Public License, version~3, through its public Git repository~\cite{OPALX}. The maintained manual provides complete worked inputs and links to the regression configurations~\cite{OPALXManual}. A reproducible calculation specifies the code revision, input and supporting field-map or distribution files, random seeds, particle number, mesh, solver and field-update settings, time-step schedule, and output configuration. Compiler and dependency versions, build options, numerical precision, hardware, execution backend, and MPI and thread counts are also relevant, particularly for performance comparisons. Reproducibility is assessed through physical observables and stated numerical tolerances, since changes in parallel reduction order can prevent bitwise identity across architectures.

\section{Beam generation and accelerator components} \label{sec:components}

\subsection{Initial distributions and emission}
The macroparticle ensemble of Sec.~\ref{sec:physics} is created from three input objects. A \texttt{BEAM} defines the particle species, the reference energy, the bunch charge, and the macroparticle allocation. A \texttt{DISTRIBUTION} describes the shape of the phase-space density to be sampled, or names a file from which particle coordinates are read. An \texttt{EMISSIONSOURCE} binds a distribution to a position and momentum offset, a start time, and an emission model. One or more sources are collected in an \texttt{EMISSIONSOURCELIST} and attached to the beam. This separation allows the same distribution shape to be injected at different positions or times, and several sources with different shapes to feed the same particle container. The complete attribute lists are given in the \OPALX{} manual~\cite{OPALXManual}; this section summarizes the underlying models and conventions.

\subsubsection{Particle species and reference particle}
The species $s$ is selected by name (electron, positron, muon, pion, proton, antiproton, deuteron, H$^-$, H$_2^+$, alpha, carbon, xenon, and uranium) or specified through an explicit rest mass $m_s$ and charge $q_s$. The input charge is specified as $q_s/e$, where $e$ is the elementary charge. The named species also carry the magnetic anomaly $G$ used by the spin tracking and, for muons and pions, the mean lifetime used by the decay processes. For generated distributions, the reference state is specified through one of the equivalent quantities $\gamma_0$, total energy $\gamma_0m_sc^2$, or momentum $\gamma_0\beta_0m_sc$, from which the reference normalized momentum $u_0=\gamma_0\beta_0$ follows. File-based distributions instead derive the reference momentum from the loaded coordinates. The reference particle follows a tracked trajectory and is distinct from the co-moving frame of Sec.~\ref{sec:self-field}, whose velocity $\boldsymbol{\beta}_b$ is derived from the ensemble at each collective update.

The bunch-charge magnitude $Q_b$ and the number of allocated macroparticles
$N_{\mathrm{alloc}}$ fix the statistical weight
\begin{equation}
    w=\frac{Q_b}{N_{\mathrm{alloc}}\,\lvert q_s\rvert},
    \qquad
    q_i = w\,q_s,
    \qquad
    m_i = w\,m_s,
    \label{eq:macro-weight}
\end{equation}
so that every macroparticle of a beam carries the charge $q_i$ and mass $m_i$ while preserving the charge-to-mass ratio $q_s/m_s$ that enters the equations of motion. Since $w$ is determined by $N_{\mathrm{alloc}}$ rather than by the number of particles actually created, creating all allocated macroparticles gives a total charge magnitude $Q_b$, with its sign determined by $q_s$. \OPALX{} warns when the requested particle numbers are inconsistent with the allocation. Charge and mass are stored as a single value per container by default, or as per-particle attributes when individual weights are required. Several beams, each with its own species and container, can be tracked simultaneously. A beam may name a daughter beam that receives the secondary particles produced by a decay process. Decay processes require one physical particle per macroparticle, $w=1$. A muon beam may carry an initial polarization vector $\bm{P}$ that enables the spin attribute of Sec.~\ref{sec:physics}.

\subsubsection{Supported distributions}
Table~\ref{tab:distributions} lists the distribution types available in \OPALX{}~\cite{OPALXManual}. The generated spatial distributions use scale parameters $\sigma_x,\sigma_y,\sigma_z$ in meters. The Gaussian types also use dimensionless normalized-momentum widths $\sigma_{u_x},\sigma_{u_y},\sigma_{u_z}$, following the convention of Sec.~\ref{sec:physics}. For Gaussian sampling, these parameters specify the widths of the underlying normal distributions; truncation and sample-mean subtraction affect the realized rms widths. For \texttt{UNIFORM}, the spatial parameters are the ellipsoid semi-axes; for the flat-top types, $\sigma_x$ and $\sigma_y$ are the transverse semi-axes, while the longitudinal profile is specified in time. Sampling is performed in local source coordinates. For generated injected distributions, the reference momentum is added, $u_z\rightarrow u_z+u_0$, followed by the source offsets $\mathbf{r}_0$ and $\mathbf{u}_0$. For generated emitted distributions, the momentum is set by the emission model described below. The number of macroparticles $N$ is specified per distribution. To represent the full specified bunch-charge magnitude, the requested particle numbers summed over all sources of a beam must equal $N_{\mathrm{alloc}}$.
\begin{table}[htb]
    \caption{\label{tab:distributions}Distribution types in \OPALX{}.
    Injected distributions create all particles at the start of the
    simulation, or at the source start time $t_0$. Emitted distributions
    release particles over a finite time interval from the plane $z=z_0$.}
    \begin{ruledtabular}
    \begin{tabular}{lll}
        Type & Mode & Phase-space model \\
        \hline
        \texttt{UNIFORM} & injected & cold, uniform within an ellipsoid \\
        \texttt{GAUSS} & injected & uncorrelated truncated Gaussian \\
        \texttt{MULTIVARIATEGAUSS} & injected & correlated truncated $6$D Gaussian \\
        \texttt{FLATTOP} & emitted & uniform disk, flat-top in time \\
        \texttt{OPALFLATTOP} & emitted & as above, \OPAL{} birth-time ordering \\
        \texttt{FROMFILE} & injected & $(x,u_x,y,u_y,z,u_z)$ from ASCII file \\
        \texttt{EMITTEDFROMFILE} & emitted & \OPAL{} emission dump with birth times \\
    \end{tabular}
    \end{ruledtabular}
\end{table}

The \texttt{UNIFORM} distribution fills the ellipsoid with semi-axes $(\sigma_x,\sigma_y,\sigma_z)$ uniformly by volume and assigns the common momentum $(0,0,u_0)$; it is intended for cold-beam test problems. The \texttt{GAUSS} distribution draws each position coordinate independently from a normal distribution truncated at $\pm3\sigma$ by inverse-transform sampling, and each momentum coordinate from an untruncated normal distribution. The \texttt{MULTIVARIATEGAUSS} distribution accepts the upper triangle of a symmetric positive-definite correlation matrix $\mathsf{C}$ in the coordinate order $(x,u_x,y,u_y,z,u_z)$. The nominal covariance matrix $\Sigma_{kl}=C_{kl}\sigma_k\sigma_l$ is factorized as $\Sigma=\mathsf{L}\mathsf{L}^{\mathsf{T}}$, and samples are formed as $\mathbf{x}=\mathsf{L}\boldsymbol{\xi}$ from standard normal variates $\boldsymbol{\xi}$ truncated at $\pm3$. Thus $\Sigma$ describes the covariance before truncation rather than the exact covariance of the generated ensemble. For both Gaussian types, the global mean of the sampled positions is subtracted, using a reduction over all MPI ranks, before the reference momentum and offsets are added; \texttt{MULTIVARIATEGAUSS} also centers the momenta. The position centroid therefore coincides with the specified source position, up to floating-point roundoff, independently of the finite particle number.

\subsubsection{Sampling and parallel generation}
Random particle sampling is performed by \Kokkos{} kernels on the selected backend using a \Kokkos{} XorShift64 random-number pool~\cite{Kokkos}. Each MPI rank generates its share of the $N$ requested macroparticles using streams seeded from the global seed option and the rank index. The local random draws require no communication, whereas the Gaussian centering operations described above require global reductions. The resulting ensemble can depend on the number of ranks and the parallel execution schedule.

For the \texttt{OPALFLATTOP} and \texttt{EMITTEDFROMFILE} emitters, a global, sorted inventory of birth times is constructed identically on every rank, and contiguous portions are assigned to individual ranks. The birth-time sequence is therefore independent of the rank count, facilitating comparisons with \OPAL{} and between parallel configurations. This property concerns the birth times and does not, by itself, establish reproducibility of the full phase-space ensemble.

\subsubsection{Loading distributions from files}
The \texttt{FROMFILE} type reads an ASCII table whose first non-comment line gives the number of records and whose next non-comment line names the columns. The required column names are \texttt{x}, \texttt{y}, \texttt{z}, \texttt{px}, \texttt{py}, and \texttt{pz}, in arbitrary order. Positions are given in meters, and the momentum columns contain the dimensionless normalized components $u_x,u_y,u_z$. Additional columns are ignored, and no unit conversion is performed. Every MPI rank reads the file and extracts a contiguous slice whose offset is obtained from an exclusive scan of the per-rank particle counts. Changing the rank count therefore changes the partition of the selected records rather than the loaded phase-space ensemble. The source offsets $\mathbf{r}_0$ and $\mathbf{u}_0$ are added after loading. With a positive start time $t_0$, the distribution is injected in the first step whose interval contains $t_0$, and the new particles receive the remaining fraction of that step, as described by Eq.~\eqref{eq:emission-dt}.

The \texttt{EMITTEDFROMFILE} type reads the legacy positional format $(x,u_x,y,u_y,t,u_z)$ of \OPAL{} emission dumps and releases the particles through time-dependent emission. Here $t$ is a pre-emission time in seconds rather than a longitudinal position. These file times are converted to tracker birth times using the legacy emission-time convention; the supported file formats and time conversions are documented in the \OPALX{} manual~\cite{OPALXManual}.

\subsubsection{Time-dependent emission}
Photo-injector simulations start from a cathode where the bunch does not yet exist as a spatial distribution. The emitted distribution types therefore describe the beam by a temporal current profile $f(t)$ and a transverse density on the source plane $z=z_0$, and particles are created step by step while the simulation time advances. Let $t_a$ denote the pulse-start time on the tracker clock and $\tau=t-t_a$ the time measured from the beginning of the pulse. Transverse positions are sampled uniformly on the elliptical disk with semi-axes $\sigma_x$ and $\sigma_y$. The longitudinal profile is a flat top of duration $t_{\mathrm{flat}}$ with Gaussian rise and fall,
\begin{equation}
    f(\tau)=
    \begin{cases}
        \exp\!\left[-\dfrac{(\tau-t_{\mathrm{r}})^2}
                              {2\sigma_{\mathrm{r}}^2}\right],
            & 0\le\tau<t_{\mathrm{r}},\\[1.2ex]
        1,
            & t_{\mathrm{r}}\le\tau
              <t_{\mathrm{r}}+t_{\mathrm{flat}},\\[0.6ex]
        \exp\!\left[
            -\dfrac{(\tau-t_{\mathrm{r}}-t_{\mathrm{flat}})^2}
                   {2\sigma_{\mathrm{f}}^2}\right],
            & t_{\mathrm{r}}+t_{\mathrm{flat}}
              \le\tau<T_{\mathrm{e}},\\[1.2ex]
        0, & \text{otherwise}.
    \end{cases}
    \label{eq:flattop-profile}
\end{equation}
where the rise and fall durations are $t_{\mathrm{r}}=c_t\sigma_{\mathrm{r}}$ and $t_{\mathrm{f}}=c_t\sigma_{\mathrm{f}}$, respectively, for longitudinal cutoff $c_t$, and the total emission time is $T_{\mathrm{e}}=t_{\mathrm{r}}+t_{\mathrm{flat}}+t_{\mathrm{f}}$. The flat-top duration follows from the specified full width at half maximum, $t_{\mathrm{flat}}=\max\{0,\,T_{\mathrm{FWHM}}-\sqrt{2\ln2}\ (\sigma_{\mathrm{r}}+\sigma_{\mathrm{f}})\}$. The flank widths are given either directly or through $10\%$--$90\%$ rise and fall times, $\sigma=T_{10\text{--}90}/\bigl(\sqrt{2\ln10}-\sqrt{2\ln(10/9)}\bigr)$. For the unmodulated profile, the normalization integral is
\begin{equation}
    A=\int_0^{T_{\mathrm{e}}}f(\tau)\,\mathrm{d}\tau
     =t_{\mathrm{flat}}
      +\sqrt{\frac{\pi}{2}}\,
       (\sigma_{\mathrm{r}}+\sigma_{\mathrm{f}})
       \,\mathrm{erf}\!\left(\frac{c_t}{\sqrt{2}}\right).
\end{equation}
The \texttt{OPALFLATTOP} type additionally supports a sinusoidal modulation of the flat top with prescribed relative amplitude and number of periods.

Emission is advanced over tracker intervals $[t,t+\Delta t]$. For emitter types that support \texttt{EMISSIONSTEPS}, the user-defined number of emission steps specifies a preferred resolution $T_{\mathrm{e}}/N_{\mathrm{steps}}$ over the emission interval~\cite{OPALXManual}. The \texttt{FLATTOP} sampler releases $\lfloor (N/A)\int_t^{t+\Delta t} f(t'-t_a)\,\mathrm{d}t'\rfloor$ new particles per step and assigns each a birth time uniformly distributed in $[t,t+\Delta t)$. The \texttt{OPALFLATTOP} sampler instead draws all $N$ birth times $t_{\mathrm{b}}$ from $f$ in advance, sorts them, and releases in each step the particles with $t_{\mathrm{b}}\le t+\Delta t$; the pulse is centered on the source start time $t_0$, and the simulation clock starts at the earliest birth time. In both cases a newly created particle $i$ receives the individual time increment
\begin{equation}
    \Delta t_i = t+\Delta t-t_{\mathrm{b},i}\in[0,\Delta t],
    \qquad
    \mathbf{r}_i \rightarrow \mathbf{r}_i+\frac{c\,\Delta t_i}{2}\frac{\mathbf{u}_i}{\gamma_i},
    \label{eq:emission-dt}
\end{equation}
which is used by the Boris integrator for the remainder of the step and compensates the half drift already performed by the existing particles (Sec.~\ref{sec:architecture}). During charge deposition, the contribution of particle $i$ is weighted by $q_i\Delta t_i/\Delta t$, so that particles born late in a step deposit a correspondingly smaller charge. Emission occurs after the space charge update of the current step, and the self-field solve is skipped entirely until the number of particles exceeds a configurable minimum; the longitudinal mesh extent is stretched in proportion to the emitted fraction while emission is active.

Two emission models define the momentum of a particle leaving the cathode. The kinetic emission energy $E_{\mathrm{kin}}$ is converted to the thermal normalized momentum $u_{\mathrm{th}}=\sqrt{(1+E_{\mathrm{kin}}/m_sc^2)^2-1}$. In the \texttt{NONE} model, $u_{\mathrm{th}}$ is added to the longitudinal component of the source momentum offset, reproducing the cathode energy increment of \OPAL{}. In the \texttt{ASTRA} model, following the treatment in the ASTRA code~\cite{Floettmann:ASTRA}, the momentum is distributed isotropically over the forward hemisphere,
\begin{equation}
    \mathbf{u}_i=\mathbf{u}_0+u_{\mathrm{th}}
    \bigl(\sin\varphi\cos\theta,\ \sin\varphi\sin\theta,\ \lvert\cos\varphi\rvert\bigr),
\end{equation}
with $\varphi=2\arccos\sqrt{\xi_1}$ and $\theta=2\pi\xi_2$ for uniform variates $\xi_1,\xi_2\in[0,1)$, which gives a uniform distribution of $\cos\varphi$ and determines the thermal emittance of the emitted beam. The non-equilibrium emission model of \OPAL{} has not been migrated. Injected distributions accept only the \texttt{NONE} model.

The source plane can also represent a conducting cathode. An emission source can request Dirichlet conditions at $z=z_0$, which is the cathode plane $z_p$ of Sec.~\ref{sec:solvers:boundary-conditions}, either through explicit image charges mirrored at the plane or through the shifted Green function of the open-boundary solver. Particles that move backward through the source plane are removed from the ensemble before the external fields are evaluated, which models their absorption by the cathode.

\subsubsection{Reference-particle initialization}
At the start of tracking, the reference particle of a container that already holds particles is set to the ensemble means $\langle\mathbf{r}_i\rangle$ and $\langle\mathbf{u}_i\rangle$. For an emitting source whose container is still empty, the reference particle is placed at the source position $\mathbf{r}_0$ with the source momentum $\mathbf{u}_0$, augmented by $\tfrac12u_{\mathrm{th}}\hat{\mathbf{z}}$ in the \texttt{ASTRA} model, and for \texttt{EMITTEDFROMFILE} with the mean momentum of the file records. The reference state is then transformed from the beamline frame to the laboratory frame and propagated to the requested start position along the lattice before the \orbitt{} of Sec.~\ref{sec:architecture} constructs the reference path. Emission itself takes place in the source frame, which is aligned with the local beamline direction at $\mathbf{r}_0$. The restrictions on checkpoint restart during emission are stated in Sec.~\ref{sec:architecture}.

\subsection{Element placement}
\OPALX{} provides accelerator components representing magnetic elements, radio-frequency structures, apertures, diagnostics, and analytical or field-map-based external fields. This section focuses on the common procedures used to place these components in the accelerator lattice and evaluate their fields. A complete list of the available elements, their parameters, and the corresponding input syntax is provided in the \OPALX{} user manual~\cite{OPALXManual}. 

Beamline elements can be placed either by specifying their entrance coordinates $s$ along a nominal design path through \texttt{ELEMEDGE}, or by specifying explicit positions and orientations. With path-based placement, the lattice geometry determines the element orientations; \texttt{PSI} can additionally specify a roll about the local longitudinal axis. Explicit placement uses three position coordinates and three orientation angles.

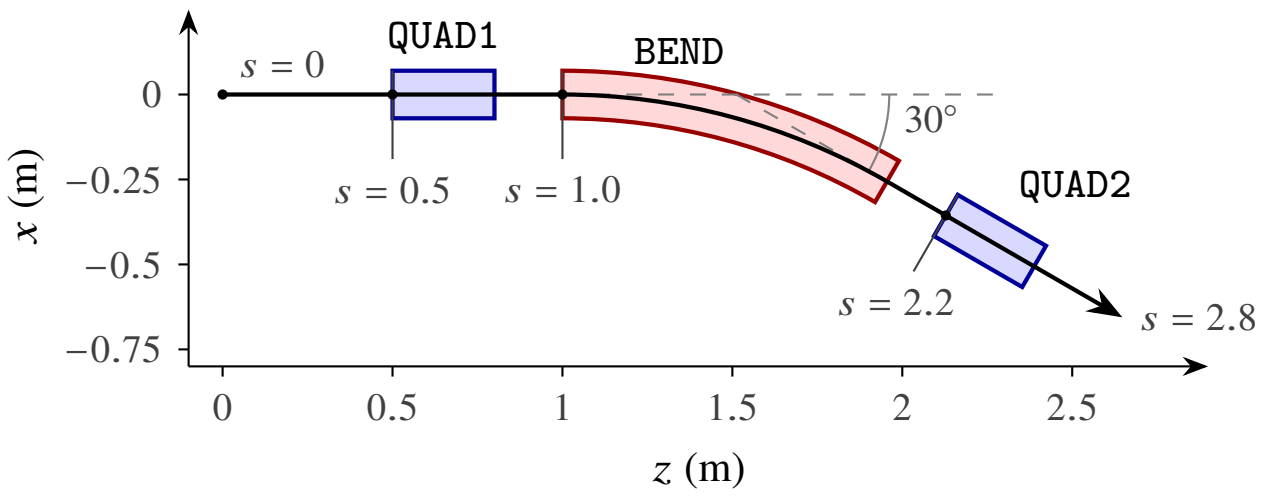
\begin{figure}[h]
  \centering
  \begin{tikzpicture}[x=0.26\columnwidth, y=0.26\columnwidth, >=Stealth,
      quad/.style   = {draw=blue!60!black, fill=blue!15, thick},
      dipole/.style = {draw=red!60!black, fill=red!15, thick},
      orbit/.style  = {thick},
      guide/.style  = {gray, dashed, thin},
      lbl/.style    = {font=\footnotesize},
      slbl/.style   = {font=\scriptsize, gray!50!black},
      entry/.style  = {gray!50!black, thin}]

    \def\qbqW{0.07}                               
    \pgfmathsetmacro{\qbqR}{6/pi}                 
    \pgfmathsetmacro{\qbqExZ}{1 + \qbqR*sin(30)}  
    \pgfmathsetmacro{\qbqExX}{-\qbqR*(1 - cos(30))} 
    \pgfmathsetmacro{\qbqTz}{1 + \qbqR*tan(15)}   
    \coordinate (C) at (1, -\qbqR);               

    \draw[->] (-0.1, -0.8) -- (2.9, -0.8);
    \node[lbl, below=13pt] at (1.4, -0.8) {$z$ (m)};
    \foreach \z in {0, 0.5, ..., 2.5}
      \draw (\z, -0.8) -- ++(0, -0.03) node[below, slbl] {$\z$};
    \draw[->] (-0.1, -0.8) -- (-0.1, 0.25);
    \node[lbl, rotate=90, anchor=south] at ([xshift=-24pt] -0.1, -0.3) {$x$ (m)};
    \foreach \x in {-0.75, -0.5, -0.25, 0}
      \draw (-0.1, \x) -- ++(-0.03, 0) node[left, slbl] {$\x$};

    \draw[quad] (0.5, -\qbqW) rectangle (0.8, \qbqW);
    \node[lbl, above] at (0.65, \qbqW) {\texttt{QUAD1}};

    \filldraw[dipole] ($(C) + (90:\qbqR+\qbqW)$)
        arc[start angle=90, end angle=60, radius=\qbqR+\qbqW]
      -- ($(C) + (60:\qbqR-\qbqW)$)
        arc[start angle=60, end angle=90, radius=\qbqR-\qbqW] -- cycle;
    \node[lbl, above] at ($(C) + (80:\qbqR+\qbqW)$) {\texttt{BEND}};

    \draw[guide] (1, 0) -- (2.3, 0);
    \draw[guide] (\qbqTz, 0) -- (\qbqExZ, \qbqExX);
    \draw[gray] (\qbqTz, 0) ++(0.45, 0) arc[start angle=0, end angle=-30, radius=0.45]
      node[pos=0.3, right, slbl] {$30^\circ$};

    \draw[orbit] (0, 0) -- (1, 0) arc[start angle=90, end angle=60, radius=\qbqR];
    \fill (0, 0) circle[radius=0.015] node[above right, slbl] {$s = 0$};

    \foreach \s in {0.5, 1.0} {
      \draw[entry] (\s, \qbqW) -- (\s, -\qbqW-0.12) node[below, slbl] {$s = \s$};
      \fill (\s, 0) circle[radius=0.015];
    }

    \begin{scope}[shift={(\qbqExZ, \qbqExX)}, rotate=-30]
      \draw[quad] (0.2, -\qbqW) rectangle (0.5, \qbqW);
      \node[lbl, above right] at (0.35, \qbqW) {\texttt{QUAD2}};
      \draw[orbit, ->] (0, 0) -- (0.8, 0) node[right, slbl] {$s = 2.8$};
      \draw[entry] (0.2, \qbqW) -- (0.2, -\qbqW-0.12) node[anchor=60, slbl] {$s = 2.2$};
      \fill (0.2, 0) circle[radius=0.015];
    \end{scope}
  \end{tikzpicture}
  \caption{Top view of the \texttt{QUAD1--BEND--QUAD2} line.
    The nominal design path~(black) is shown to $s=2.8$~m.
    The gray lines mark the entrance faces and give their
    \texttt{ELEMEDGE} path length $s$.
    The magnet widths are not to scale.}
  \label{fig:qbq_layout}
\end{figure}

These two placement syntaxes are demonstrated below using the same three-element lattice, consisting of a quadrupole of length $30\,\si{\centi\metre}$, a sector bend of length $1\,\si{\metre}$ with a bending angle of $30^\circ$, and a second quadrupole of length $30\,\si{\centi\metre}$. The elements are denoted \verb|QUAD1|, \verb|BEND|, and \verb|QUAD2|, respectively. The nominal geometry is shown in Fig.~\ref{fig:qbq_layout}. For both examples, $s=0$ is at the laboratory origin and the initial design direction is along the positive $z$ axis.

\subsubsection{Placement Along Path Length}
The \texttt{ELEMEDGE} attribute specifies the entrance coordinate $s$, and \texttt{L} specifies the nominal element length. For the sector bend \texttt{SBEND}, \texttt{L} is the arc length of the nominal design path.
\begin{verbatim}
    QUAD1: QUADRUPOLE, 
        ELEMEDGE = 0.5,
        L = 0.3,    
        K1 =  1.0;
        
    BEND: SBEND,
        ELEMEDGE = 1.0,
        L = 1.0, 
        ANGLE = PI / 6, 
        HGAP = 0.0;
        
    QUAD2: QUADRUPOLE, 
        ELEMEDGE = 2.2,
        L = 0.3,    
        K1 = -1.0;

    QBQ: LINE = (QUAD1, BEND, QUAD2);
\end{verbatim}

\subsubsection{Absolute Placement}
For the quadrupoles and sector bend in this example, \texttt{X}, \texttt{Y}, and \texttt{Z} specify the entrance-frame origin in laboratory coordinates. The orientation angles \texttt{THETA}, \texttt{PHI}, and \texttt{PSI} correspond to rotations about the $y$, $x$, and $z$ axes, respectively, and are specified in radians.
\begin{verbatim}
    REAL X_QUAD2 = -(6 / PI) * (1 - COS(PI / 6))
                   - 0.2 * SIN(PI / 6);
    REAL Z_QUAD2 = 1.0 + (6 / PI) * SIN(PI / 6)
                   + 0.2 * COS(PI / 6);

    QUAD1: QUADRUPOLE,
        X = 0.0, Y = 0.0, Z = 0.5,
        THETA = 0.0, PHI = 0.0, PSI = 0.0,
        L = 0.3,
        K1 =  1.0;

    BEND: SBEND,
        X = 0.0, Y = 0.0, Z = 1.0,
        THETA = 0.0, PHI = 0.0, PSI = 0.0,
        L = 1.0,
        ANGLE = PI / 6,
        HGAP = 0.0;

    QUAD2: QUADRUPOLE,
        X = X_QUAD2, Y = 0.0, Z = Z_QUAD2,
        THETA = -PI / 6, PHI = 0.0, PSI = 0.0,
        L = 0.3,
        K1 = -1.0;

    QBQ: LINE = (QUAD1, BEND, QUAD2);
\end{verbatim}
The bend radius is $\rho=L_B/\theta=(6/\pi)\,\si{\metre}$, where $L_B=1\,\si{\metre}$ and $\theta=\pi/6$. The $0.2\,\si{\metre}$ drift from the bend exit to the second quadrupole gives the expressions for \texttt{X\_QUAD2} and \texttt{Z\_QUAD2} above. Setting \texttt{THETA=-PI/6} aligns the second quadrupole with the nominal bend-exit direction. The two inputs therefore describe the same nominal geometry.

\subsection{Element Field Evaluation}
The element-selection procedures described in Sec.~\ref{sec:architecture} identify candidate contributions to the external fields. For tracking based on a reference path, the \orbitt{} constructs an \texttt{IndexMap} that associates path-length intervals with candidate elements according to their field-support extents~\cite{OPALXManual}. For each particle, its position is transformed into the relevant element-local coordinates, and the electric and magnetic fields are evaluated using the local field model or field map. The local field-support checks determine whether a candidate element contributes to that particle. The field vectors are transformed back to the tracker coordinates before summation. The assembled external fields are combined with the self-fields as described in Sec.~\ref{sec:particle-field} and used in the Boris momentum update. The element-dependent time sampling is specified in Sec.~\ref{sec:time-integration}.

\begin{figure}[h]
  \centering
  \begin{tikzpicture}[x=0.2\columnwidth, y=0.155\columnwidth, >=Stealth,
      rf/.style      = {draw=red!60!black, fill=red!15, thick},
      sol/.style     = {draw=blue!60!black, fill=blue!15, thick},
      rfline/.style  = {red!60!black, line width=0.3pt},
      solline/.style = {blue!60!black, line width=0.6pt},
      guide/.style   = {gray, dashed, thin},
      lbl/.style     = {font=\footnotesize},
      slbl/.style    = {font=\scriptsize, gray!50!black},
      set/.style     = {font=\scriptsize\ttfamily, gray!50!black, rotate=90, anchor=west}]

    \def\h{0.3}                                     
    \def\rowRF{2.19} \def\rowA{1.77} \def\rowB{1.35} 
    \def\solenoid{plot[smooth] coordinates {(0,0.040) (0.25,0.043) (0.286,0.048) (0.321,0.059)
      (0.357,0.093) (0.393,0.155) (0.429,0.207) (0.464,0.234) (0.5,0.247) (0.536,0.253)
      (0.571,0.256) (0.607,0.258) (0.643,0.259) (0.679,0.260) (0.714,0.260) (0.75,0.260)
      (0.786,0.260) (0.821,0.260) (0.857,0.259) (0.893,0.258) (0.929,0.256) (0.964,0.253)
      (1,0.247) (1.04,0.234) (1.07,0.207) (1.11,0.155) (1.14,0.093) (1.18,0.059) (1.21,0.048)
      (1.25,0.043) (1.5,0.040)}}

    \foreach \s in {3.3, 4.2, 4.8, 5.05, 5.7, 5.9, 6.55, 7.4}
      \draw[guide] (\s, 0) -- (\s, \rowRF+\h);

    \draw[rf] (3.3, \rowRF) rectangle (7.398, \rowRF+\h);
    \draw[rfline, shift={(0, \rowRF)}] plot[domain=3.3:7.398, samples=1200]
      (\x, {0.15 + 0.12*0.5*(tanh(max(-4, min(4, (\x-3.345)/0.01)))
                            - tanh(max(-4, min(4, (\x-7.353)/0.01))))
                 *sin(360*(\x-3.3)/0.09993)});
    \fill[red!15, path fading=west] (6.85, \rowRF) rectangle (7.05, \rowRF+\h);
    \fill[red!15] (7.05, \rowRF) rectangle (7.398, \rowRF+\h);
    \draw[rf, fill=none] (3.3, \rowRF) rectangle (7.398, \rowRF+\h);
    \node[lbl, anchor=north east, inner sep=1.5pt] at (7.398, \rowRF+\h) {\texttt{TW1}};

    \foreach \e/\row/\name in {3.3/\rowA/S1, 4.2/\rowB/S2, 5.05/\rowA/S3, 5.9/\rowB/S4} {
      \draw[sol] (\e, \row) rectangle (\e+1.5, \row+\h);
      \node[lbl, anchor=north east, inner sep=1.5pt] at (\e+1.5, \row+\h) {\texttt{\name}};
      \draw[solline, shift={(\e, \row)}] \solenoid;
    }

    \foreach \s/\set in {3.75/{TW1, S1}, 4.925/{TW1, S2}, 5.375/{TW1, S2, S3},
                         5.8/{TW1, S3}, 6.225/{TW1, S3, S4}, 6.975/{TW1, S4}}
      \node[set] at (\s, 0.05) {\set};
    \node[set, black] at (4.6, 0.05) {TW1, S1, S2};   

    \draw[->] (3.15, 0) -- (7.7, 0);
    \node[lbl, below=13pt] at (5.4, 0) {$s$ (m)};
    \foreach \s in {3.5, 4.0, 4.5, 5.0, 5.5, 6.0, 6.5, 7.0, 7.5}
      \draw (\s, 0) -- ++(0, -0.05) node[below, slbl] {$\s$};

    \draw (4.33, 0) -- (4.33, \rowRF+0.5) node[above, lbl] {$s^\ast$};
    \fill (4.33, 0) circle[radius=1.2pt];
  \end{tikzpicture}
\caption{Illustration of overlapping field supports in the first SwissFEL linac section along the reference-path coordinate $s$. \texttt{TW1} is a traveling-wave structure, and \texttt{S1}--\texttt{S4} are solenoids. The curves schematically indicate the on-axis RF electric field $E_z$ at a fixed time and the solenoid magnetic fields $B_z$. The dashed lines mark the intervals of the \texttt{IndexMap} constructed by the \orbitt{}. At $s^\ast$, the candidate elements are \texttt{TW1}, \texttt{S1}, and \texttt{S2}; their contributions are summed wherever the particle lies within their local field supports.}
  \label{fig:field_overlap}
\end{figure}
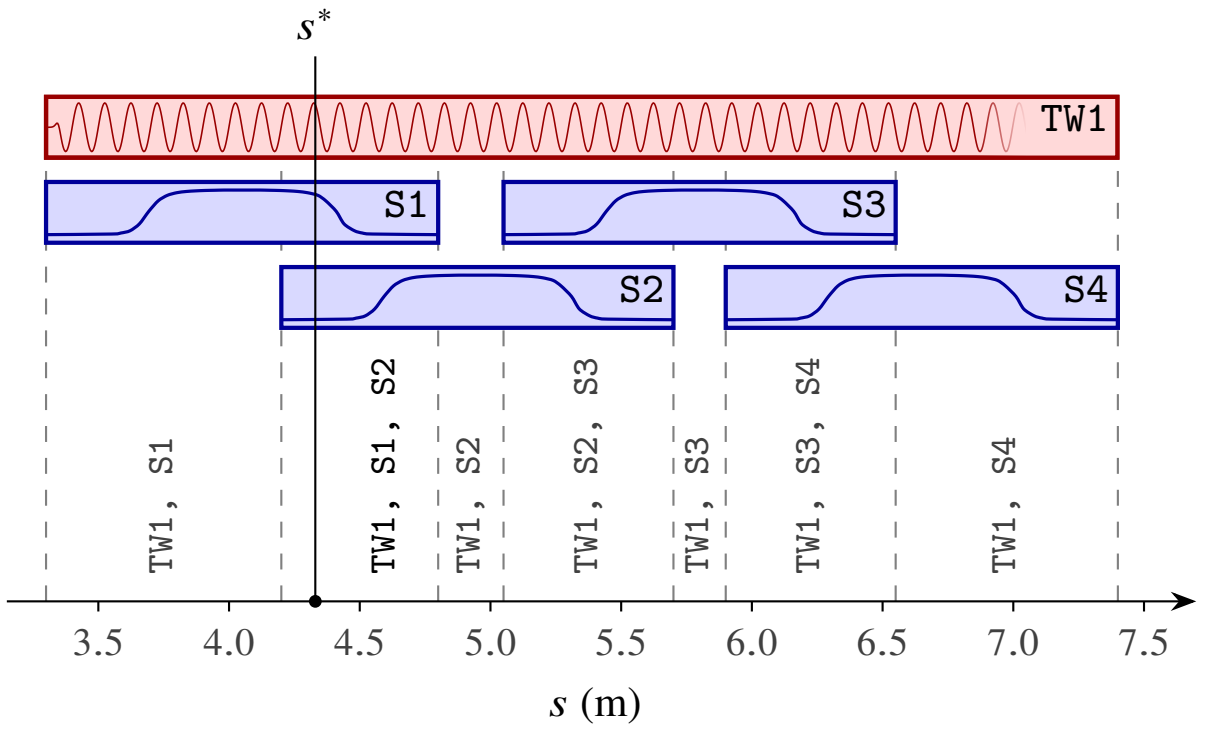

Figure~\ref{fig:field_overlap} illustrates overlapping external-field supports in the first SwissFEL linac section. At the marked path coordinate $s^\ast$, the \texttt{IndexMap} identifies the traveling-wave structure \texttt{TW1} and the solenoids \texttt{S1} and \texttt{S2}. Their field contributions are summed for particles within the corresponding local field supports.

\section{Numerical verification and cross-code benchmarking}
\label{sec:validation}

\subsection{Numerical verification}

\subsubsection{Disorder-Induced Heating}
The disorder-induced heating~(DIH) test follows the confined spherical-bunch benchmark of~\textcite{Mitchell:2015}. The initial distribution is cold and uniform within a sphere of radius $R=1.774\,\si{\milli\metre}$, with bunch-charge magnitude $Q_b=25\,\si{\nano\coulomb}$. The \texttt{CONSTANTFOCUSING} element supplies a linear focusing field whose strength is initialized from the initial mean space-charge field and then held fixed to suppress bulk expansion. The initial rms size is $\sigma_x=R/\sqrt{5}=0.793\,\si{\milli\metre}$.

For the electron macroparticle weight $w=10^6$, the numerical particles have charge magnitude $\lvert q_i\rvert=we$ and mass $M=wm_e$, consistently with Eq.~\eqref{eq:macro-weight}. Let $N$ denote the number of numerical particles and $n=3N/(4\pi R^3)$ their initial number density. For the stationary confined system, the corresponding plasma period is
\begin{equation}
    T_p=2\pi\sqrt{\frac{\epsilon_0M}{nq_i^2}}
       \simeq43.1\,\si{\pico\second}.
\end{equation}
As correlations develop, the correlation potential energy decreases and thermal kinetic energy increases. For an initially cold, confined one-component plasma, the equilibrium coupling parameter is approximately $\Gamma_{\mathrm{eq}}=2.23$~\cite{Maxson:2013, Mitchell:2015}. Defining the Wigner--Seitz radius $a_{\mathrm{WS}}=[3/(4\pi n)]^{1/3}$, the equilibrium-temperature estimate is
\begin{equation}
    k_BT_{\mathrm{eq}}
    \simeq\frac{q_i^2}
    {4\pi\epsilon_0a_{\mathrm{WS}}\Gamma_{\mathrm{eq}}},
\end{equation}
where $k_B$ is the Boltzmann constant. For an isotropic, nonrelativistic thermal distribution without position--momentum correlations, using the initial rms size gives the equilibrium-emittance estimate
\begin{equation}
    \varepsilon_{n,x}^{\mathrm{eq}}
    \simeq\sigma_x\sqrt{\frac{k_BT_{\mathrm{eq}}}{Mc^2}}
    =\frac{R}{\sqrt{5}}\sqrt{\frac{k_BT_{\mathrm{eq}}}{Mc^2}}
    \simeq4.91\,\si{\micro\metre}.
\end{equation}
This estimate describes the simulated macroparticle system. Its interpretation is subject to the macroparticle-weight limitation discussed in Sec.~\ref{sec:p3m}.

\begin{figure}[ht]
    \centering
    \includegraphics[width=\columnwidth]{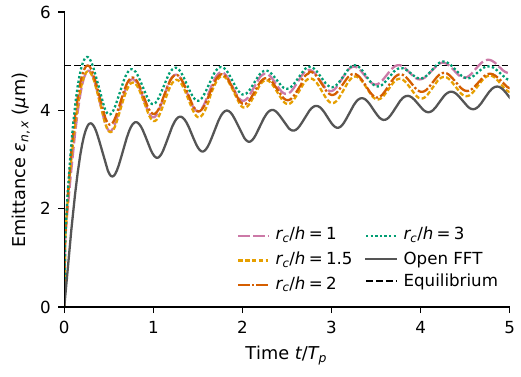}
    \caption{Disorder-induced heating of an initially cold, uniformly populated sphere with radius $R=1.774\,\si{\milli\metre}$ and macroparticle weight $w=10^6$. All runs use open field boundaries and a $128^3$ mesh. The P3M curves correspond to $r_c/h=1$, $1.5$, $2$, and $3$, where $h$ is the mesh spacing; the open FFT result is shown for comparison. The normalized emittance $\varepsilon_{n,x}$ is plotted against $t/T_p$. The dashed line indicates the equilibrium estimate.}
    \label{fig:p3m-disorder-induced-heating-128-grid-ranks-8-emittance-no-0p5h}
\end{figure}

Figure~\ref{fig:p3m-disorder-induced-heating-128-grid-ranks-8-emittance-no-0p5h} compares the P3M results for $r_c/h=1$, $1.5$, $2$, and $3$ against the open-boundary mesh solver of Sec.~\ref{sec:open-solver}. The P3M results show rapid initial emittance growth followed by oscillations, with peaks near the equilibrium estimate. The open FFT result remains below the P3M curves over the displayed five plasma periods. The tested P3M cutoffs produce similar oscillatory evolution, illustrating the role of the short-range interactions. This comparison is specific to the stated configuration and does not establish a general cutoff criterion.

\subsubsection{Frozen Spin Setup}
For planar motion in a transverse magnetic field, the muon's magnetic anomaly produces spin precession relative to the momentum, known as anomalous or $g$--$2$ precession~\cite{jegerlehner_muon_2009}. An appropriately chosen radial electric field cancels this relative precession. The resulting frozen-spin technique is the basis of the muEDM search at PSI~\cite{adelmann_compact_2025,farley_new_2004}.

Let $\theta$ denote the orbital azimuth and $\alpha$ the signed spin phase relative to the momentum in the orbital plane, with both angles using the same azimuthal orientation. For circular motion with constant speed and fields perpendicular to the velocity, the phase advance per unit orbital angle is
\begin{equation}
    \frac{d\alpha}{d\theta}=\gamma\,\frac{GB-\big(\frac{1}{\beta^2\gamma^2}-G\big)\beta E/c}{B+E/(\beta c)}.
    \label{eq:phase-advance}
\end{equation}
Here $G=(g-2)/2$ is the magnetic anomaly, $\beta=v/c$, and $\gamma=(1-\beta^2)^{-1/2}$. The quantities $B$ and $E$ are the magnetic-field magnitude and the radial electric-field magnitude, respectively, evaluated on the orbit. The signs in Eq.~\eqref{eq:phase-advance} correspond to an inward radial electric field, with the electric and magnetic Lorentz forces both directed toward the orbit center. Setting the numerator to zero gives
\begin{equation}
    E_f=\frac{GBc\,\beta\gamma^2}{1-G\beta^2\gamma^2}.
\end{equation}
At $E=E_f$, the magnetic-moment contribution to the spin precession relative to the momentum vanishes in this ideal configuration.

\begin{figure}[h]
    \centering
    \includegraphics[width=0.99\linewidth]{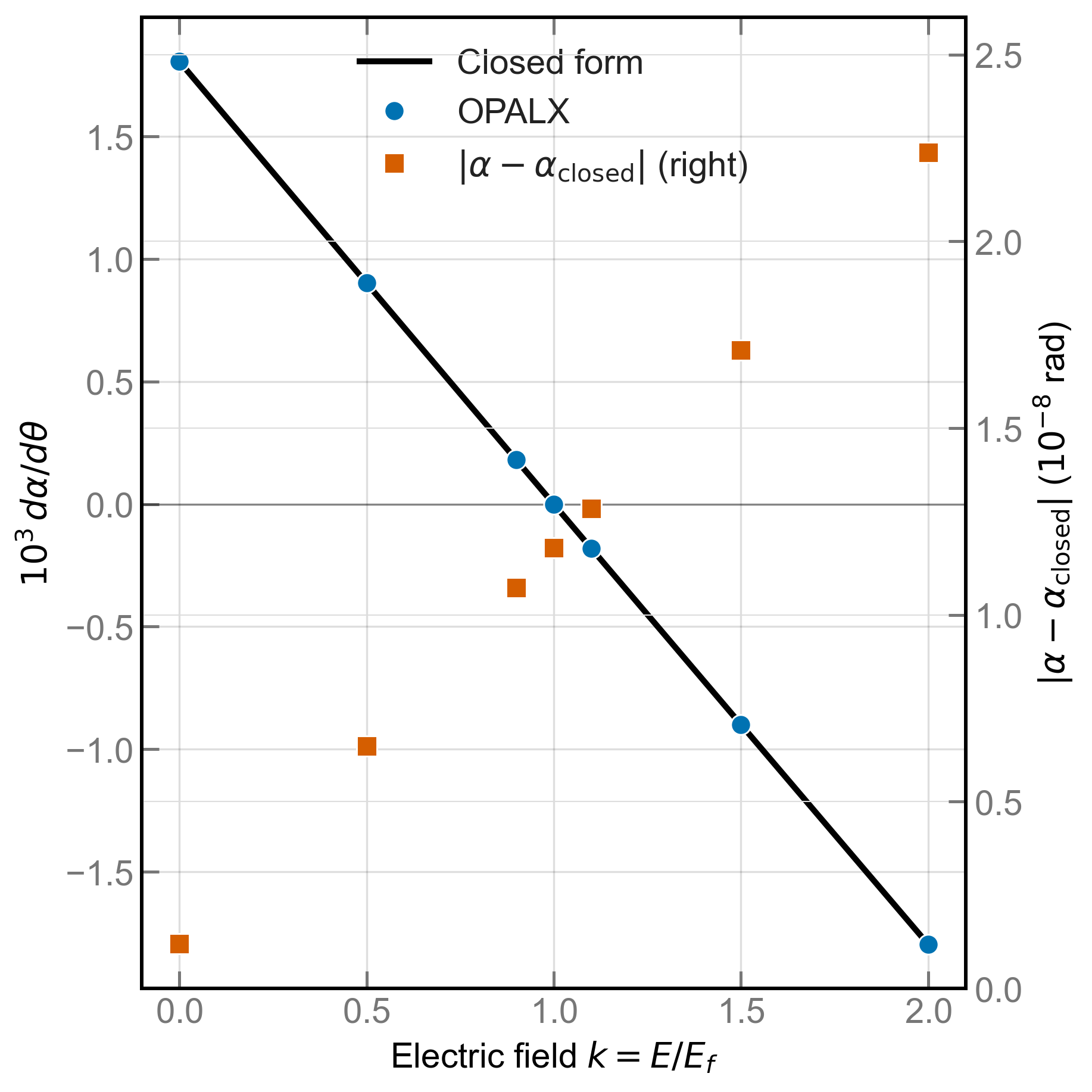}
    \caption{Spin phase advance per unit orbital angle, $d\alpha/d\theta$, as a function of the electric-field scale $k=E/E_f$ (left axis). The black curve is the analytical reference from Eq.~\eqref{eq:phase-advance}, and blue circles show the \OPALX{} estimates. Orange squares show the absolute differences from the analytical prediction using the right-axis scale. The phase advance vanishes at the frozen-spin setting $k=1$.}
    \label{fig:frozen-spin}
\end{figure}

The numerical test uses a $\mu^+$ with momentum $p=125\,\si{\mega\electronvolt}/c$ ($\gamma=1.5491$) and initial polarization aligned with its momentum. A three-dimensional field map represents a uniform vertical magnetic field of $B=3\,\si{\tesla}$ and the radial electric field of a cylindrical capacitor. The particle is tracked with a time step $\Delta t=0.1\,\si{\pico\second}$, and its polarization and momentum are recorded after an orbital arc of approximately $30^\circ$. Figure~\ref{fig:frozen-spin} compares the numerical phase-advance estimates with Eq.~\eqref{eq:phase-advance} as the electric-field scale $k=E/E_f$ is varied. The phase advance changes sign near $k=1$, where the frozen-spin condition is satisfied.

\subsection{Cross-code benchmarking}

\subsubsection{\OPALX{}--TraceWin beam-dynamics comparison}
\label{sec:opalx-tracewin}

We compare the beam-dynamics predictions of \OPALX{} with those of \trw{}~\cite{UriotTraceWin:2015} for a low-energy ion beam transported through a quadrupole triplet followed by an interdigital H-mode drift-tube linac (IH-DTL) cavity. The cavity electromagnetic fields are represented by a three-dimensional field map. This configuration tests beam transport through magnetic focusing elements and an RF accelerating structure.

\begin{figure}[tb]
    \centering
    \includegraphics[width=\columnwidth]
        {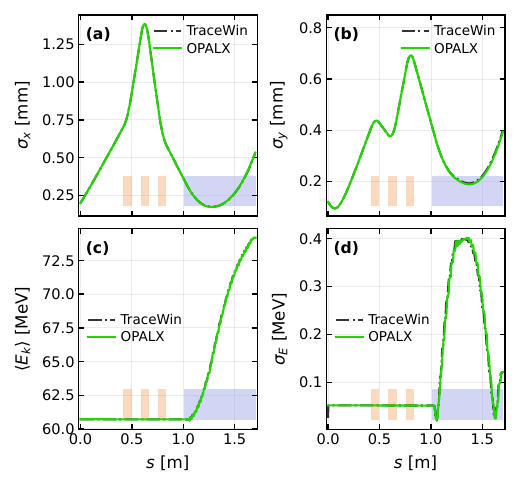}
    \caption{Comparison of \OPALX{} and \trw{} for an ion beam transported through a quadrupole triplet followed by an IH-DTL cavity. Panels show (a) the horizontal rms beam size, (b) the vertical rms beam size, (c) the mean kinetic energy, and (d) the rms energy spread as functions of the longitudinal coordinate $s$. Gray dash-dotted curves show \trw{} results, and green solid curves show \OPALX{} results. Orange and blue shaded regions indicate the quadrupoles and the IH-DTL cavity, respectively.}
    \label{fig:opalx-tracewin}
\end{figure}

Figure~\ref{fig:opalx-tracewin} compares the horizontal and vertical rms beam sizes, the mean kinetic energy, and the rms energy spread as functions of the longitudinal coordinate $s$. The two codes predict closely matching transverse beam envelopes, including the focusing and defocusing through the quadrupole triplet and the subsequent evolution through the cavity. The mean kinetic-energy profiles agree during both the initial transport and RF acceleration. The energy-spread profiles also show consistent evolution, including the increase within the accelerating section.

The mean absolute difference between the predicted mean kinetic energies is $0.0155\,\mathrm{MeV}$. At the end of the line, the relative differences in the horizontal and vertical rms beam sizes are each below $2\,\%$. These endpoint differences are small compared with the variations in beam size along the line. The comparison provides an independent cross-check of the predicted magnetic focusing and RF acceleration for this configuration and these beam observables.

\subsubsection{G4Beamline Wien Filter}
A Wien filter combines a transverse magnetic field $B_y$ and an orthogonal electric field $E_x$~\cite{wien_untersuchungen_1898,plies_wien_2011}. For a particle traveling along the positive $z$ axis in ideal uniform fields, the condition $E_x=vB_y$ cancels the transverse Lorentz force. The spin polarization can nevertheless rotate in the combined fields~\cite{tsuno_chapter_2013}. This configuration therefore provides a test of spin transport in simultaneous electric and magnetic fields. We compare \OPALX{} with \gfbl{}~\cite{roberts_g4beamline_2007}, a Geant4-based particle-tracking code with spin-tracking and decay capabilities~\cite{agostinelli_geant4_2003}.

The test uses the separator field map from the \gfbl{} model of the $\mu$H2 beamline at PSI. The fields are scaled to central values $B_y=22.1\,\si{\milli\tesla}$ and $E_x=1.60\,\si{\mega\volt\per\metre}$, and the input muon momentum is $p=26.825\,\si{\mega\electronvolt}/c$. Both codes use the same field map and the same set of $1000$ initial $\mu^+$ particles. Positions, momenta, and spin vectors are recorded at eight monitor planes.

\begin{figure}[h]
    \centering
    \includegraphics[width=0.99\linewidth]{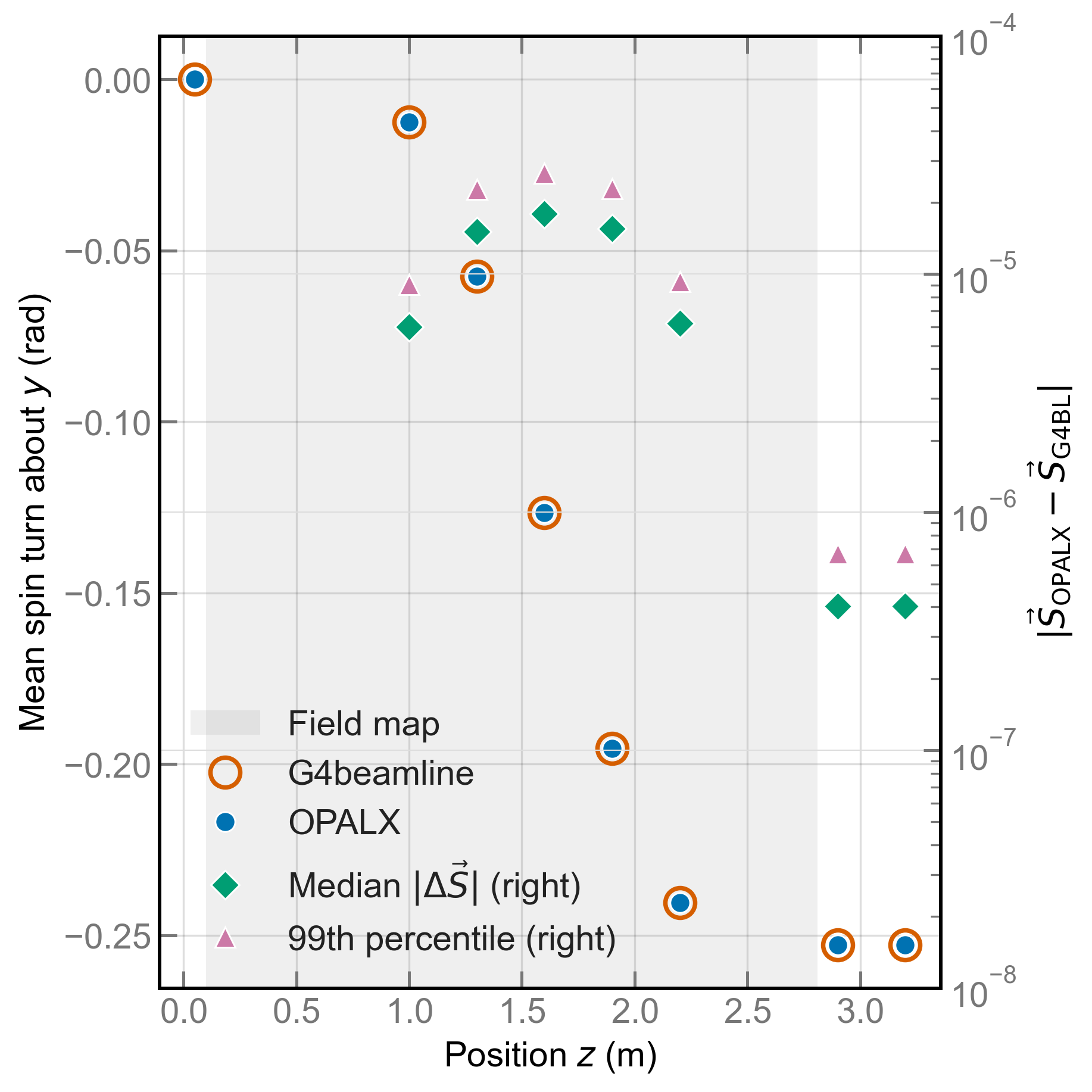}
    \caption{Mean spin rotation about the $y$ axis as a function of position $z$ (left axis). Blue filled circles show \OPALX{} results, and orange open circles show \gfbl{} results. The shaded region indicates the longitudinal extent of the field map. Green diamonds and purple triangles show the median and $99$th percentile, respectively, of the spin-vector differences (right axis, logarithmic scale). The initial difference is zero to machine precision and is omitted from the logarithmic axis.}
    \label{fig:wien-turn}
\end{figure}

Figure~\ref{fig:wien-turn} compares the mean spin rotation about the $y$ axis along the separator and summarizes the spin-vector differences between the two codes using the median and $99$th percentile. Both codes predict a downstream mean rotation of approximately $-0.25\,\si{\radian}$.

\subsubsection{G4Beamline \texorpdfstring{$\mu$E4}{muE4} Beamline}
\label{sec:mue4-comparison}
The $\mu$E4 beamline at PSI transports surface muons~\cite{prokscha_new_2008}. These $\mu^+$ originate from the decay at rest of $\pi^+$ that stop near the surface of a graphite target irradiated by the $590$~MeV proton beam. The comparison presented here uses an input muon momentum of $28$~MeV/$c$.

The modeled beamline segment is $19.25$~m long and contains one solenoid (WSX), three dipoles (one ASR61 and two ASR62), twelve quadrupoles (QSM600) arranged in four triplets, and three sets of collimator slits (FS61, FS62, and FS63). The magnetic fields are specified by field maps. Material interactions are not simulated: particles entering the collimators or magnet housings are removed. Twenty-three monitor planes record the transverse phase space and spin polarization along the beamline. The initial spin polarization is aligned with the beam axis.

For the comparison, \OPALX{} and \gfbl{} track the same set of $2\times10^4$ initial muons. \OPALX{} uses a time step of $10^{-12}$~s, corresponding at $28$~MeV/$c$ to the $0.0768$~mm step length used in \gfbl{}. Two cases are considered, both with spin tracking: one with muon decay disabled and one with muon decay enabled. Particle data are recorded at all 23 monitor planes.

\begin{figure}[h]
    \centering
    \includegraphics[width=0.99\linewidth]{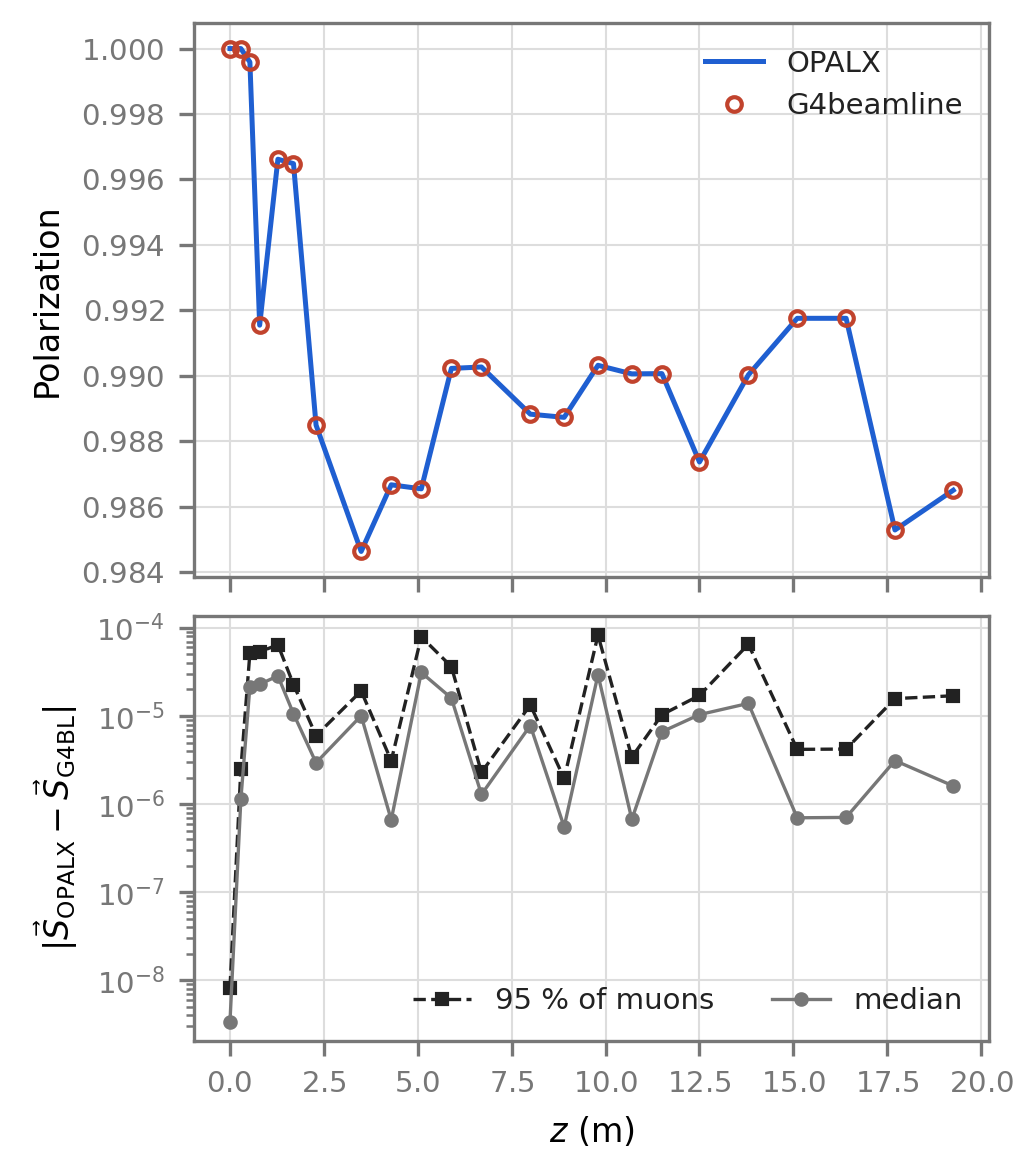}
    \caption{Spin-polarization comparison along the $\mu$E4 beamline. Upper panel: magnitude of the mean spin-polarization vector from \OPALX{} and \gfbl{}. Lower panel: median and 95th percentile of the magnitudes of the particlewise differences between the spin-polarization vectors recorded by the two codes.}
    \label{fig:mue4-polarization}
\end{figure}

With decay disabled, the histogram bin populations at the monitors agree to within one particle. For 95\,\% of the muons, the position differences are within $12~\mu$m and the spin-vector differences are within $10^{-4}$. Figure~\ref{fig:mue4-polarization} compares the magnitude of the mean spin polarization and the particlewise spin differences along the beamline.

\begin{figure*}[t]
    \centering
    \captionsetup[subfigure]{justification=centering}
    \begin{subfigure}[t]{0.32\textwidth}
        \centering
        \includegraphics[width=\linewidth]{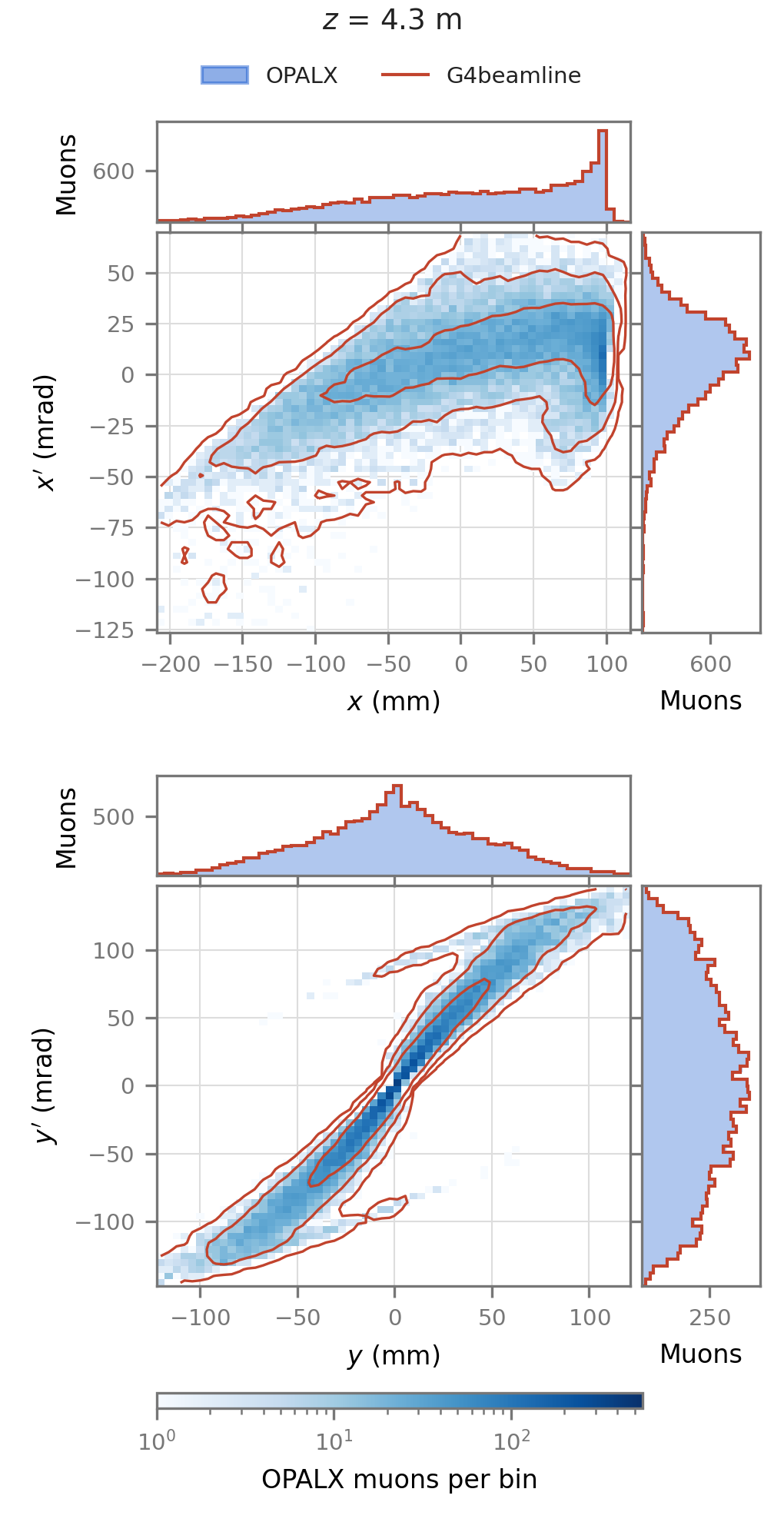}
        \caption{}
        \label{fig:mue4-transverse-4300}
    \end{subfigure}\hfill
    \begin{subfigure}[t]{0.32\textwidth}
        \centering
        \includegraphics[width=\linewidth]{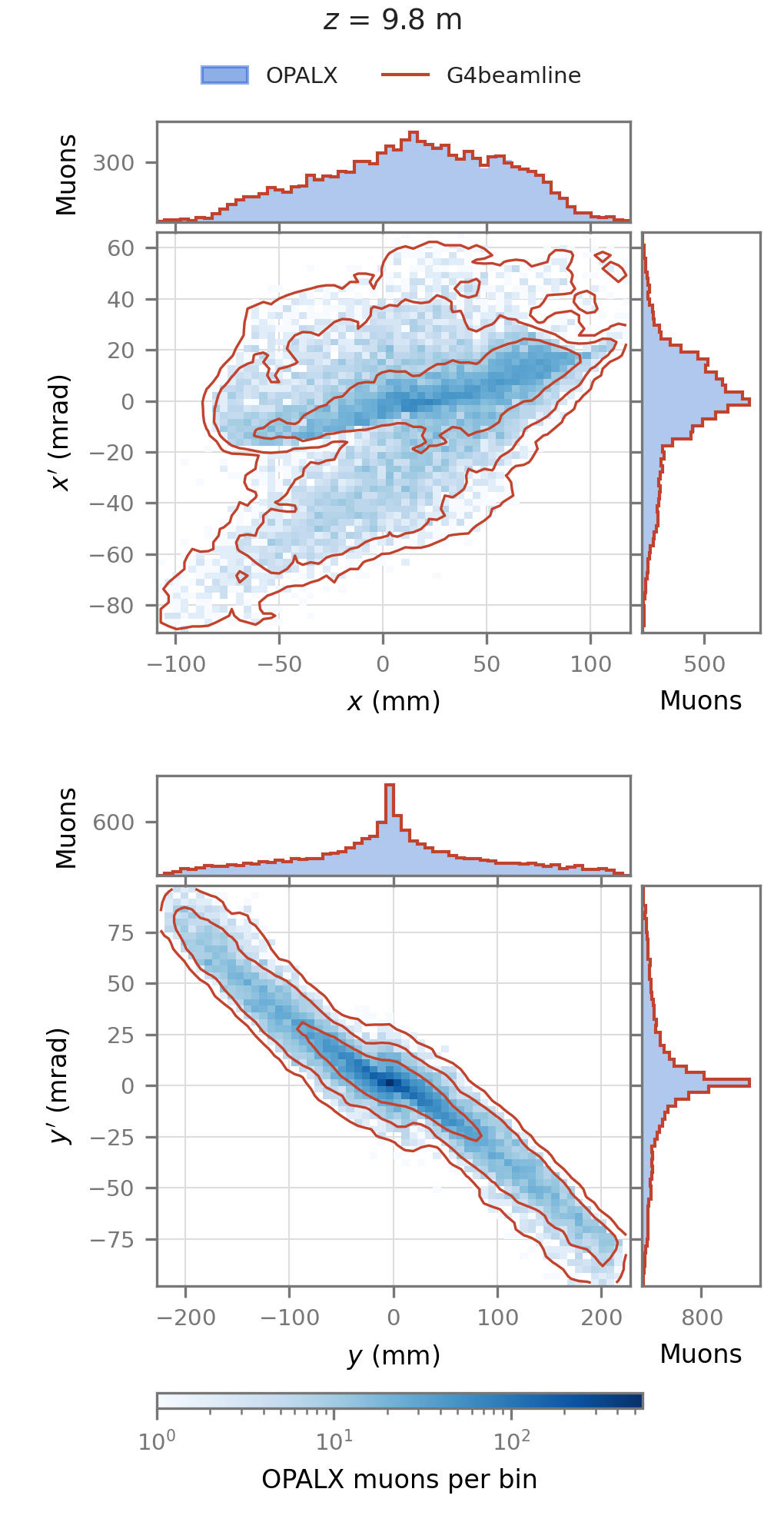}
        \caption{}
        \label{fig:mue4-transverse-9800}
    \end{subfigure}\hfill
    \begin{subfigure}[t]{0.32\textwidth}
        \centering
        \includegraphics[width=\linewidth]{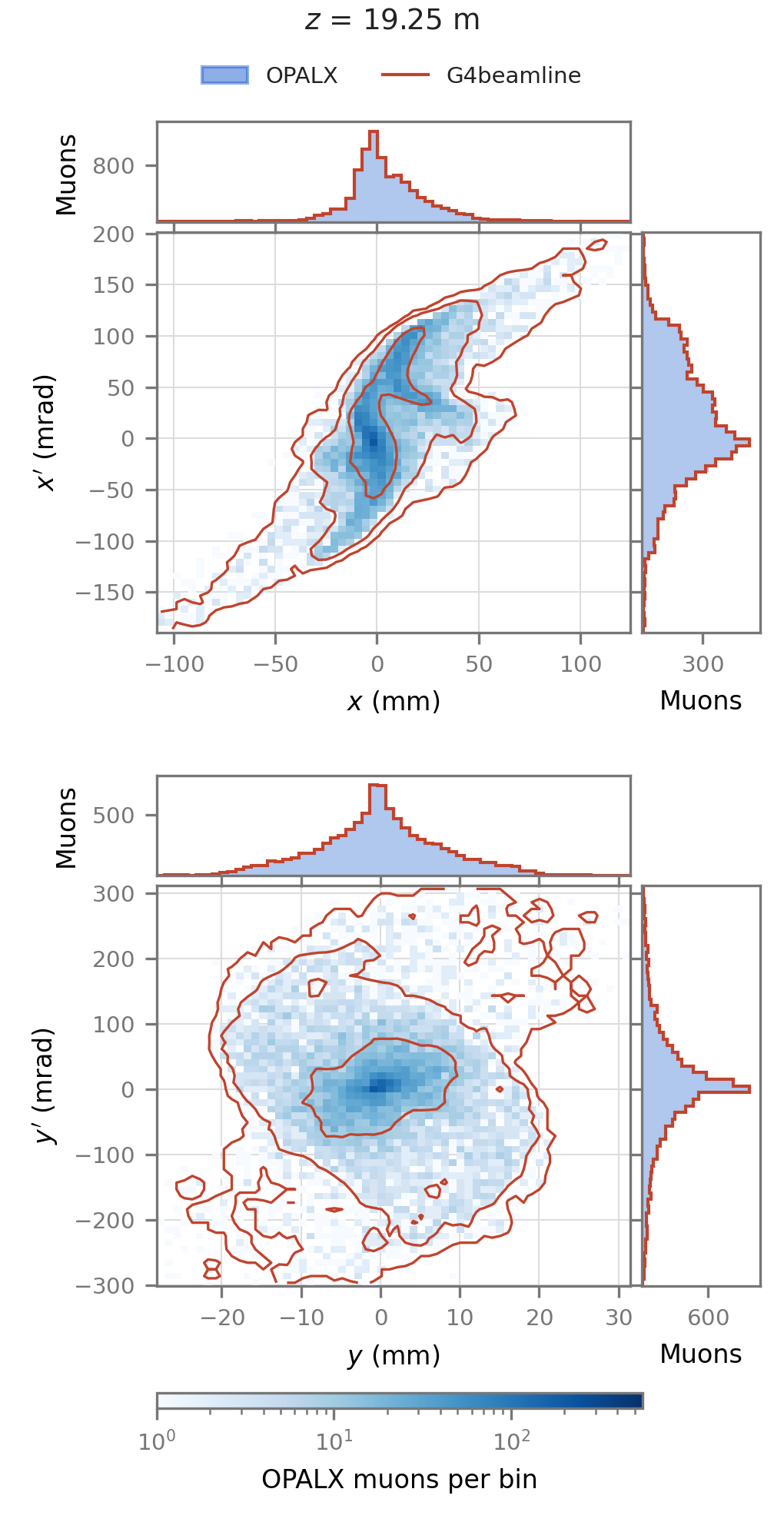}
        \caption{}
        \label{fig:mue4-transverse-19250}
    \end{subfigure}
    \caption{Transverse phase-space distributions in the $\mu$E4 comparison with spin tracking and muon decay enabled, using $2\times10^4$ initial muons. The monitor locations are (a) $z=4.3$\,m, after the first dipole and collimator, (b) $z=9.8$\,m, and (c) $z=19.25$\,m, at the end of the modeled segment. The upper and lower panels show the horizontal and vertical phase spaces, respectively. Blue shading shows the \OPALX{} particle counts per bin; red contours enclose 50\,\%, 90\,\%, and 99\,\% of the \gfbl{} particles. The marginal histograms show the projections onto each axis.}
    \label{fig:mue4-transverse}
\end{figure*}

With decay enabled, the independently sampled decays require a statistical comparison of the two ensembles. Approximately 11\,\% of the muons decay before the end of the line. The final monitor records 9262 muons in \OPALX{} and 9283 in \gfbl{}; particles can also be lost at the collimators and magnet housings. Figure~\ref{fig:mue4-transverse} compares the transverse phase-space distributions at three monitor locations for this case.

\section{Performance portability and scalability}
\label{sec:performance}

\subsection{Three-Dimensional Space Charge Benchmark}

We assess the particle throughput of \OPALX{} using the three-dimensional space charge benchmark of Ref.~\cite{Huebl:2026}. The reference configuration is a $1\,\si{\nano\coulomb}$, $250\,\si{\mega\electronvolt}$ electron bunch transported through a $6\,\si{\metre}$ drift with one space charge kick, using a $64^3$ mesh, open boundaries, and an integrated Green function. The calculations use double precision (FP64). The benchmark is formulated for codes that advance particles along a longitudinal coordinate $s$; the \OPALX{} input adapts this configuration to laboratory-time tracking. All reported \OPALX{} points pass the benchmark's physics checks.

\begin{figure}[tb]
    \centering
    \includegraphics[width=\columnwidth]{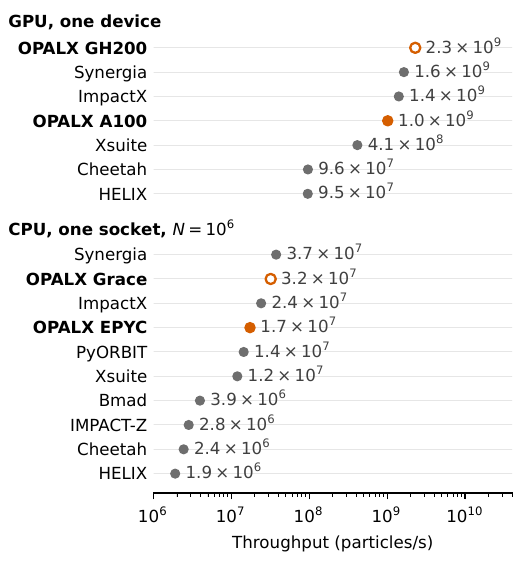}
    \caption{Throughput summary for the three-dimensional space charge benchmark of Fig.~\ref{fig:sc3d_throughput_vs_n}. GPU entries use the particle count with the highest measured throughput for each code; CPU entries use $N=10^6$. Orange markers identify \OPALX{}: filled markers denote the A100-40GB GPU and EPYC~7742 CPU results, and open markers denote the GH200 GPU and Grace CPU results. Reference CPU results use the EPYC~7763. The ordering therefore includes different hardware configurations and, for GPUs, potentially different particle counts.}
    \label{fig:sc3d_ranking}
\end{figure}

\begin{figure*}[t]
    \centering
    \includegraphics[width=\textwidth]{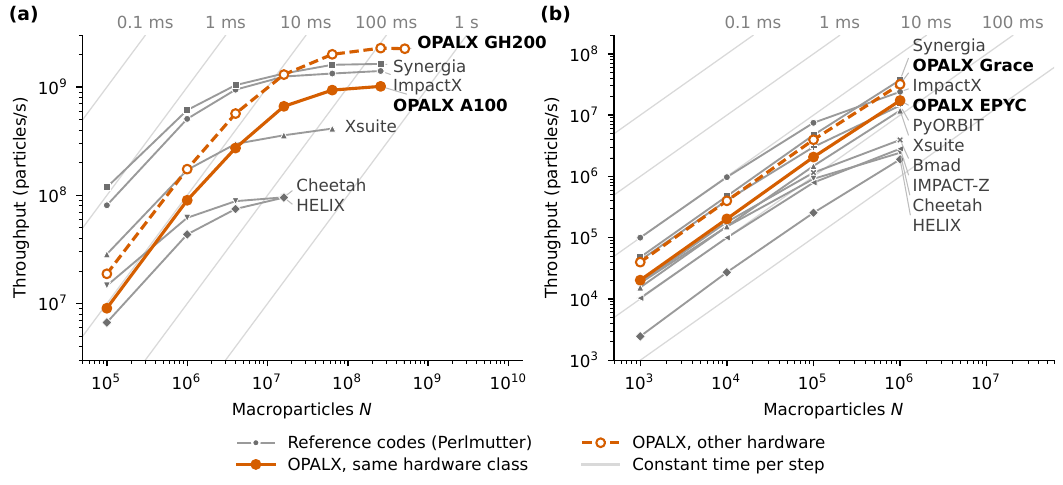}
    \caption{Throughput as a function of macroparticle count $N$ for the three-dimensional space charge benchmark~\cite{Huebl:2026}: (a) one GPU and (b) one CPU socket. The reference configuration uses a $1\,\si{\nano\coulomb}$, $250\,\si{\mega\electronvolt}$ electron bunch, a $64^3$ mesh, an integrated Green function, open boundaries, and FP64 arithmetic. Gray curves show reference results obtained with the benchmark harness on Perlmutter using an A100-SXM4-40GB GPU or a 64-core AMD EPYC~7763 socket, with the fastest FP64 build and CPU MPI-rank/thread layout selected for each code and particle count. Orange solid curves with filled markers show \OPALX{} on an A100-SXM4-40GB GPU and a 64-core AMD EPYC~7742 socket; dashed curves with open markers show \OPALX{} on a GH200 GPU and 64 Grace CPU cores. The EPYC~7742 and EPYC~7763 are different processor generations. The \OPALX{} CPU runs use one MPI rank per core. Light lines indicate constant time per step, with throughput $N/\tau$.}
    \label{fig:sc3d_throughput_vs_n}
\end{figure*}

Figure~\ref{fig:sc3d_throughput_vs_n} compares throughput as a function of the number of macroparticles $N$, and Fig.~\ref{fig:sc3d_ranking} summarizes the results. Throughput is defined as $N/\tau$, where $\tau$ is the time per step. For \OPALX{}, this time is estimated from the difference between the wall times of 11-step and 1-step runs, divided by the ten additional steps. The reported timings use the fastest of three to five repeats. This metric measures the benchmark step, including the space charge calculation, rather than the particle momentum update alone.

The reference results were obtained with the benchmark harness on Perlmutter, using an NVIDIA A100-SXM4-40GB GPU or a 64-core AMD EPYC~7763 CPU socket. The reference CPU runs use the fastest MPI-rank and thread layout selected for each code and particle count, and the reference results use the fastest FP64 build of each code. The \OPALX{} GPU runs use an A100-SXM4-40GB and an NVIDIA GH200. Its CPU runs use a 64-core AMD EPYC~7742 socket and 64 of the 72 cores of a Grace CPU, with one MPI rank per core. The A100 comparison uses the same GPU model as the reference results. The CPU comparisons involve different processor models, and the GH200 results use different GPU hardware.

On a single GPU, the \OPALX{} throughput approaches $1.0\times10^9$ macroparticles/s on the A100 and $2.3\times10^9$ macroparticles/s on the GH200. The A100 throughput is approximately 62\,\% of the Synergia result and 72\,\% of the ImpactX result on the 40~GB A100, approximately $2.5$ times the Xsuite throughput, and approximately $11$ times the Cheetah and HELIX throughputs. For $N\lesssim10^6$, the A100 step time changes little with particle count: $11.0$~ms at $N=10^5$ and $11.1$~ms at $N=10^6$. This behavior is consistent with a substantial contribution from mesh operations and other costs that depend weakly on particle count. Above approximately $N=10^7$, the throughput approaches a plateau, consistent with an increasing contribution from particle-dependent work.

At $N=10^6$, \OPALX{} reaches $3.2\times10^7$ macroparticles/s on 64 Grace cores. This is approximately $2.7$, $2.2$, and $1.3$ times the reference throughputs of Xsuite, PyORBIT, and ImpactX, respectively, and 85\,\% of the Synergia throughput on the EPYC~7763. On the EPYC~7742, \OPALX{} reaches $1.7\times10^7$ macroparticles/s, approximately 72\,\% of the ImpactX throughput, 46\,\% of the Synergia throughput, and $1.2$ and $1.5$ times the PyORBIT and Xsuite throughputs, respectively. These ratios describe the measured code--hardware combinations.

In a CPU rank-scaling test, increasing the number of MPI ranks from 8 to 64 reduces the field-solve time from $145$ to $23$~ms and the total step time from $179$ to $31$~ms. The particle-related contribution decreases from $28$ to $6$~ms. These runs use orthogonal recursive bisection (ORB) load balancing, with the load-balancing condition checked at every step. Without ORB, the particle-related contribution remains between $50$ and $58$~ms, with the ranks covering the bunch center holding most of the particles.

\subsection{\texorpdfstring{$\mu$E4}{muE4} Beamline Performance and Strong Scaling}
\label{sec:mue4-performance}
We measure run times for the $\mu$E4 beamline model described in Sec.~\ref{sec:mue4-comparison}, using larger particle ensembles with spin tracking and muon decay enabled and space charge disabled. These runs use a $28$~MeV/$c$ beam, a time step of $10^{-11}$~s in \OPALX{}, and the corresponding step length of $0.768$~mm in \gfbl{}. Recording monitors are placed near the start ($z=10$~mm) and at the end ($z=19.25$~m) of the modeled segment. Three hardware configurations are used. On the CSCS Daint supercomputer, \OPALX{} tracks $10^8$ muons sampled from the same initial distribution as the smaller comparison case, using 1--16 NVIDIA GH200 GPUs with one MPI rank per GPU. On a MacBook Pro with an Apple M4 processor, each code uses one CPU core to track the same $10^5$ muons read from a file. On the PSI Merlin6 cluster, \gfbl{} tracks $10^6$ muons on 60 CPU cores of one node.

\begin{figure*}[t]
    \centering
    \begin{subfigure}[t]{0.48\textwidth}
        \centering
        \includegraphics[width=\linewidth]{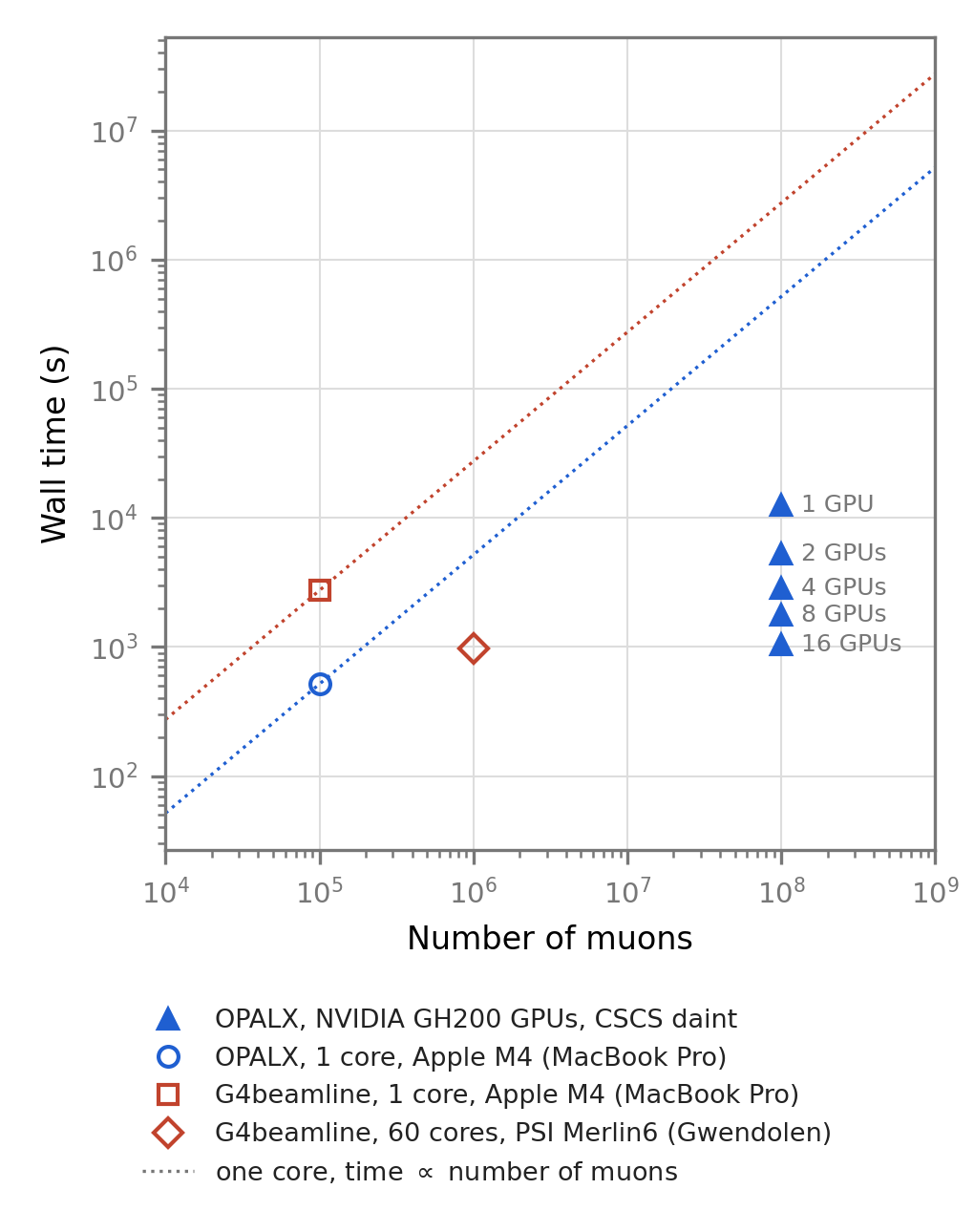}
        \caption{Wall time as a function of the initial number of muons. Each point represents one run. Dotted lines show linear extrapolations through the single-core measurements.}
        \label{fig:mue4-performance}
    \end{subfigure}\hfill
    \begin{subfigure}[t]{0.48\textwidth}
        \centering
        \includegraphics[width=\linewidth]{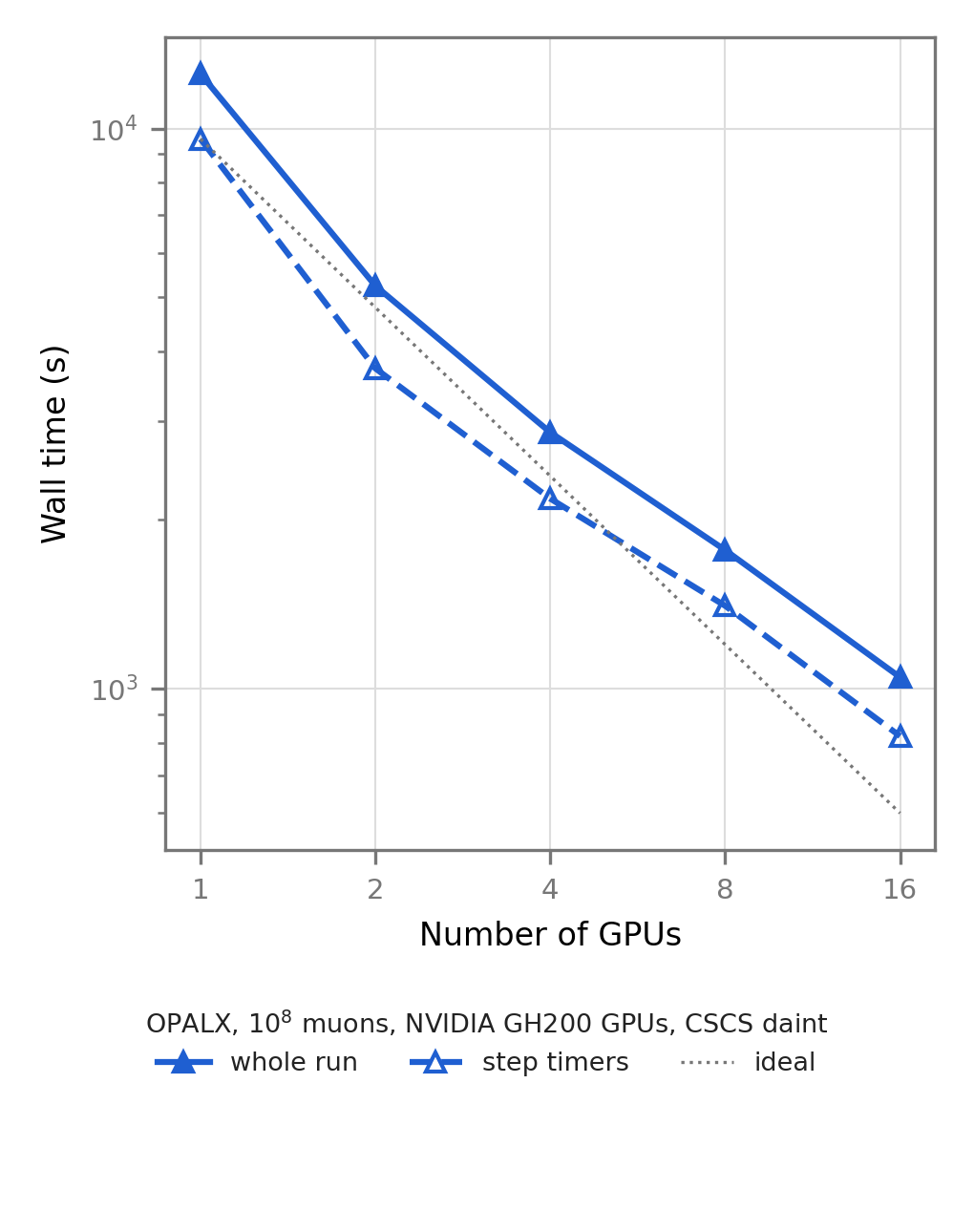}
        \caption{Strong scaling of \OPALX{} on NVIDIA GH200 GPUs for $10^8$ initial muons. Solid line: total wall time. Dashed line: accumulated step timers covering particle motion, field evaluation, aperture checks, and beam statistics. Dotted line: ideal scaling normalized to the one-GPU step-timer measurement.}
        \label{fig:mue4-strong-scaling}
    \end{subfigure}
    \caption{Run-time measurements for the $\mu$E4 model with spin tracking and muon decay enabled and space charge disabled.}
    \label{fig:mue4-scaling}
\end{figure*}

Figure~\ref{fig:mue4-performance} shows the measured wall times. On one Apple M4 core, the wall time per initial muon is $5.2$~ms for \OPALX{} and $27$~ms for \gfbl{}, giving a run-time ratio of approximately $5.2$. The dotted lines extrapolate these measurements assuming that wall time is proportional to the initial particle count. Under this assumption, tracking $10^8$ muons on one core would require approximately 6 days with \OPALX{} and 32 days with \gfbl{}. In comparison, the measured \OPALX{} wall time for $10^8$ muons is approximately $3.5$~h on one GH200 GPU and $17$~min on 16 GPUs. The one-GPU time is about 41 times shorter than the extrapolated single-core \OPALX{} time. The \gfbl{} run on 60 Merlin6 cores requires $0.98$~ms per initial muon, approximately $7.8$ times the wall time per initial muon of the one-GPU \OPALX{} run. These latter comparisons involve different hardware and, for the Merlin6 run, different particle counts.

Figure~\ref{fig:mue4-strong-scaling} shows the strong scaling of \OPALX{} for a fixed initial ensemble of $10^8$ muons. Increasing the GPU count from 1 to 16 reduces the wall time from $12\,520$~s to $1047$~s, corresponding to a speedup of $12.0$ and a parallel efficiency of approximately 75\,\%. The accumulated step timers show a speedup of $11.7$. Time outside these timers, attributed primarily to initialization, monitor output, muon decay, and removal of lost particles, accounts for 23\,\% of the total wall time on one GPU and 21\,\% on 16 GPUs. This contribution therefore also decreases in absolute time. The reduction in wall time from one to two GPUs exceeds a factor of two; the cause of this superlinear speedup has not been established.

\section{Current limitations and outlook}
\label{sec:outlook}
The results presented here assess selected physical models and numerical operations within \OPALX{}. The verification cases and cross-code comparisons in Sec.~\ref{sec:validation} examine short-range interactions, orbital and spin dynamics, and RF acceleration for specific configurations. Extending this assessment requires systematic studies of time-step, mesh, and particle-number dependence, together with uncertainty estimates for stochastic processes. Agreement between independent codes provides a useful consistency check, while comparisons with measurements are needed to assess predictive accuracy for experimental systems.

The applicability of the collective models is governed by the assumptions described in Sec.~\ref{sec:physics}. The self-field calculation uses a quasi-static electrostatic approximation in a co-moving frame, requiring sufficiently small internal currents and sufficiently slow source evolution. The local treatment of accelerating or curved beams also requires the frame approximation to remain appropriate on the bunch scale. The 2.5D method introduces additional assumptions about beam elongation and longitudinal variation and currently uses a planar reference path. For P3M, the interpretation of short-range dynamics depends on the numerical particle weights, as discussed in Sec.~\ref{sec:p3m}. Resolving pair interactions between macroparticles does not by itself establish the physical binary-collision dynamics of the represented beam. The disorder-induced-heating case therefore tests the specified numerical particle system.

The performance results in Sec.~\ref{sec:performance} demonstrate CPU and GPU execution for the reported workloads. Their interpretation depends on the particle count, numerical model, timing interval, and hardware configuration. In particular, the multi-GPU $\mu$E4 calculation has space charge disabled, so its strong-scaling result does not establish the scaling of a distributed self-consistent field calculation. Further performance assessment should include integrated accelerator workloads with collective effects, communication and load-balancing costs, and the sensitivity of time to solution to the required numerical accuracy.

Priorities for further assessment include comparisons with legacy \OPAL{} calculations and experimental measurements, together with integrated accelerator studies. Studies combining accelerator fields, collective dynamics, and, where relevant, emission or spin tracking will help establish the range of practical applications supported by the framework. Reproducible assessment requires benchmark inputs, supporting field maps and distributions, analysis scripts, and numerical results, together with the code and build information needed to reproduce both the physical observables and the performance measurements.

\section{Conclusions}
\label{sec:conclusions}

We have presented \OPALX{}, an accelerator simulation framework that integrates relativistic particle tracking, spin dynamics, accelerator-component models, and collective-field calculations within a common computational architecture. Built on \IPPL{} and \Kokkos{}, the framework combines distributed-memory parallelism with CPU and GPU execution. Its interfaces connect beam generation and emission, external-field evaluation, space charge solvers, and diagnostics within a shared tracking workflow.

The cases presented in this paper assess selected parts of this implementation. The disorder-induced-heating case examines correlation-driven heating in the specified numerical particle system, while the frozen-spin case compares spin evolution with an analytical reference. Comparisons with \trw{} assess magnetic focusing and RF acceleration, and comparisons with \gfbl{} examine orbital and spin transport through a Wien filter and the $\mu$E4 beamline. These studies provide cross-checks for the configurations and observables considered.

The performance benchmarks demonstrate CPU and GPU execution for both a three-dimensional space charge calculation and beamline tracking. In the space charge benchmark, \OPALX{} reaches approximately $1.0\times10^9$ macroparticles/s on an NVIDIA A100 GPU and $2.3\times10^9$ macroparticles/s on an NVIDIA GH200 GPU. For the $\mu$E4 model with spin tracking and muon decay enabled and space charge disabled, tracking $10^8$ initial muons on 16 GH200 GPUs gives a speedup of approximately $12$ relative to one GPU, corresponding to a parallel efficiency of 75\,\%. These results establish performance for the reported workloads and hardware configurations. The broader convergence, validation, and application studies outlined in Sec.~\ref{sec:outlook} will extend this assessment.

\begin{acknowledgments}
This work was supported in part by the Platform for Advanced Scientific Computing~(PASC) through the {\it OPAL-X: Object oriented Parallel Accelerator Library for eXascale} project and by the Swiss National Science Foundation~(SNSF). 
\end{acknowledgments}


\newpage
\bibliography{references}

@PREAMBLE{
 "\providecommand{\noopsort}[1]{}" 
 # "\providecommand{\singleletter}[1]{#1}%" 
}

@misc{OPAL,
  author        = {Adelmann, Andreas and Calvo, Pedro and Frey, Matthias 
                   and Gsell, Achim and Locans, Uldis and Metzger-Kraus, Christof
                   and Neveu, Nicole and Rogers, Chris and Russell, Steve
                   and Sheehy, Suzanne and others},
  title         = {{OPAL} a Versatile Tool for Charged Particle Accelerator
                   Simulations},
  year          = {2019},
  eprint        = {1905.06654},
  archivePrefix = {arXiv},
  primaryClass  = {physics.acc-ph},
  doi           = {10.48550/arXiv.1905.06654},
  url           = {https://arxiv.org/abs/1905.06654}}

@misc{OPALX,
  author        = {{OPALX Project}},
  title         = {{OPALX}: Object-Oriented Particle Accelerator Library 
                   for Exascale},
  year          = {2026},
  url           = {https://github.com/OPALX-project/OPALX},
  howpublished  = {\url{https://github.com/OPALX-project/OPALX}}}

@manual{OPALXManual,
  author        = {{OPALX Project}},
  title         = {The {OPALX} Universe: User Guide, Physics Manual, Reference, 
                   and Developer Handbook},
  year          = {2026},
  url           = {https://opalx-project.github.io/opalx-manual/},
  note          = {Source: \url{https://github.com/OPALX-project/opalx-manual}}}

@inproceedings{IPPL,
  author        = {Muralikrishnan, Sriramkrishnan and Frey, Matthias
                   and Vinciguerra, Alessandro and Ligotino, Michael
                   and Cerfon, Antoine J. and Stoyanov, Miroslav
                   and Gayatri, Rahulkumar and Adelmann, Andreas},
  title         = {Scaling and Performance Portability of the Particle-in-Cell
                   Scheme for Plasma Physics Applications through Mini-Apps
                   Targeting Exascale Architectures},
  booktitle     = {Proceedings of the 2024 {SIAM} Conference on Parallel
                   Processing for Scientific Computing ({PP})},
  pages         = {26--38},
  year          = {2024},
  publisher     = {Society for Industrial and Applied Mathematics},
  doi           = {10.1137/1.9781611977967.3}}

@article{Kokkos,
  author        = {Trott, Christian R. and Lebrun-Grandi{\'e}, Damien
                   and Arndt, Daniel and Ciesko, Jan and Dang, Vinh
                   and Ellingwood, Nathan and Gayatri, Rahulkumar
                   and Harvey, Evan and Hollman, Daisy S. and Ibanez, Dan
                   and others},
  title         = {{Kokkos} 3: Programming Model Extensions for the Exascale Era},
  journal       = {{IEEE} Transactions on Parallel and Distributed Systems},
  volume        = {33},
  number        = {4},
  pages         = {805--817},
  year          = {2022},
  doi           = {10.1109/TPDS.2021.3097283}}

@manual{MPI,
  author        = {{Message Passing Interface Forum}},
  title         = {{MPI}: A Message-Passing Interface Standard, Version 5.0},
  year          = {2025},
  url           = {https://www.mpi-forum.org/docs/mpi-5.0/mpi50-report.pdf}}

@inProceedings{heFFTe,
  author        = "Ayala, Alan and Tomov, Stanimire and Haidar, Azzam and Dongarra, Jack",
  title         = {{heFFTe: Highly Efficient FFT for Exascale}},
  booktitle     = "Computational Science - {ICCS} 2020",
  series        = {Lecture Notes in Computer Science},
  volume        = {12137},
  pages         = {262--275},
  publisher     = {Springer International Publishing},
  year          = {2020},
  doi           = {10.1007/978-3-030-50371-0_19}}

@article{Zampetakis:2024,
  author        = {Zampetakis, M. and Antoniou, F. and Asvesta, F.
                   and Bartosik, H. and Papaphilippou, Y.},
  title         = {Interplay of space charge, intrabeam scattering,
                   and synchrotron radiation in the {Compact Linear
                   Collider} damping rings},
  journal       = {Phys. Rev. Accel. Beams},
  volume        = {27},
  pages         = {064403},
  year          = {2024},
  doi           = {10.1103/PhysRevAccelBeams.27.064403}}

@misc{zampetakis:2025,
  title         = {Interplay of space charge and intrabeam scattering 
                   in the LHC ion injector chain}, 
  author        = {Zampetakis, M. and Antoniou, F. and Asvesta, F. 
                   and Bartosik, H. and Papaphilippou, Y. 
                   and Sa{\'a} Hern{\'a}ndez, A.},
  year          = {2023},
  eprint        = {2310.03504},
  archivePrefix = {arXiv},
  primaryClass  = {physics.acc-ph},
  url           = {https://arxiv.org/abs/2310.03504}}

@book{Reiser:2008,
  author        = {Reiser, Martin},
  title         = {Theory and Design of Charged Particle Beams},
  edition       = {2},
  publisher     = {Wiley-VCH},
  address       = {Weinheim, Germany},
  year          = {2008},
  doi           = {10.1002/9783527622047}}

@book{Hofmann:2017,
  author        = {Hofmann, Ingo},
  title         = {Space Charge Physics for Particle Accelerators},
  series        = {Particle Acceleration and Detection},
  publisher     = {Springer International Publishing},
  address       = {Cham, Switzerland},
  year          = {2017},
  doi           = {10.1007/978-3-319-62157-9}}

@article{Qiang:2000,
  author        = {Qiang, Ji and Ryne, Robert D. and Habib, Salman
                   and Decyk, Viktor},
  title         = {An {Object-Oriented Parallel Particle-in-Cell Code}
                   for {Beam Dynamics Simulation} in {Linear Accelerators}},
  journal       = {Journal of Computational Physics},
  volume        = {163},
  number        = {2},
  pages         = {434--451},
  year          = {2000},
  doi           = {10.1006/jcph.2000.6570}}

@inproceedings{Huebl:2022,
  author        = {Huebl, A. and Leh{\'e}, R. and Mitchell, C. E.  
                   and J. Qiang and R.D. Ryne and R.T. Sandberg and J.-L. Vay},
  title         = {{Next Generation Computational Tools for the Modeling and 
                    Design of Particle Accelerators at Exascale}},
  booktitle     = {Proc. 5th Int. Particle Accel. Conf. (NAPAC'22)},
  pages         = {302--306},
  eid           = {TUYE2},
  publisher     = {JACoW Publishing},
  month         = {10},
  year          = {2022},
  doi           = {10.18429/JACoW-NAPAC2022-TUYE2}}

@article{Neveu:2019,
  author        = {Neveu, N. and Spentzouris, L. and Adelmann, A. and Ineichen, Y. 
                   and Kolano, A. and Metzger-Kraus, C. and Bekas, C. and Curioni, A. 
                   and Arbenz, P.},
  title         = {Parallel general purpose multiobjective optimization framework 
                   with application to electron beam dynamics},
  journal       = {Phys. Rev. Accel. Beams},
  volume        = {22},
  number        = {5},
  pages         = {054602},
  numpages      = {11},
  year          = {2019},
  month         = {May},
  doi           = {10.1103/PhysRevAccelBeams.22.054602}}

@article{Winklehner:2017,
  title         = {Realistic simulations of a cyclotron spiral inflector 
                   within a particle-in-cell framework},
  author        = {Winklehner, Daniel and Adelmann, Andreas and Gsell, Achim 
                   and Kaman, Tulin and Campo, Daniela},
  journal       = {Phys. Rev. Accel. Beams},
  volume        = {20},
  number        = {12},
  pages         = {124201},
  numpages      = {12},
  year          = {2017},
  doi           = {10.1103/PhysRevAccelBeams.20.124201}}

@article{Winklehner:2022,
  author        = {Winklehner, Daniel and Conrad, Janet M and Schoen, Devin 
                   and Yampolskaya, Maria and Adelmann, Andreas 
                   and Mayani, Sonali and Muralikrishnan, Sriramkrishnan},
  title         = {Order-of-magnitude beam current improvement in compact cyclotrons},
  journal       = {New Journal of Physics},
  volume        = {24},
  number        = {2},
  pages         = {023038},
  year          = {2022},
  doi           = {10.1088/1367-2630/ac5001}}

@article{Fano:1957,
  author        = {Fano, U.},
  title         = {Description of {States} in {Quantum} {Mechanics} 
                   by {Density} {Matrix} and {Operator} {Techniques}},
  journal       = {Reviews of Modern Physics},
  volume        = {29},
  number        = {1},
  pages         = {74--93},
  year          = {1957},
  doi           = {10.1103/RevModPhys.29.74}}

@book{Blum:2012,
  author        = {Blum, Karl},
  title         = {Density {Matrix} {Theory} and {Applications}},
  series        = {Springer {Series} on {Atomic}, {Optical}, 
                   and {Plasma} {Physics}},
  volume        = {64},
  edition       = {3},
  publisher     = {Springer},
  address       = {Berlin, Heidelberg},
  year          = {2012},
  doi           = {10.1007/978-3-642-20561-3}}

@article{Bargmann:1959,
  author        = {Bargmann, V. and Michel, Louis and Telegdi, V. L.},
  title         = {Precession of the Polarization of Particles Moving 
                   in a Homogeneous Electromagnetic Field},
  journal       = {Phys. Rev. Lett.},
  volume        = {2},
  number        = {10},
  pages         = {435--436},
  year          = {1959},
  doi           = {10.1103/PhysRevLett.2.435}}

@article{Thomas:1926,
  author        = {Thomas, L. H.},
  title         = {The {Motion} of the {Spinning} {Electron}},
  journal       = {Nature},
  volume        = {117},
  number        = {2945},
  pages         = {514},
  year          = {1926},
  doi           = {10.1038/117514a0}}

@book{Jackson:1999,
  author        = {Jackson, John David},
  title         = {Classical Electrodynamics},
  edition       = {3},
  publisher     = {Wiley},
  address       = {New York},
  year          = {1999}}

@article{Mayani:2025,
  author        = {Mayani, Sonali and Montanaro, Veronica 
                   and Cerfon, Antoine J. and Frey, Matthias 
                   and Muralikrishnan, Sriramkrishnan 
                   and Adelmann, Andreas},
  title         = {A Massively Parallel Performance Portable Free-Space 
                   Spectral Poisson Solver},
  journal       = {ACM Transactions on Mathematical Software},
  volume        = {51},
  number        = {3},
  eid           = {15},
  pages         = {1--23},
  year          = {2025},
  doi           = {10.1145/3748815}}

@article{Qiang:2006,
  author        = {Qiang, Ji and Lidia, Steve M. and Ryne, Robert D.
                   and Limborg-Deprey, Cecile},
  title         = {Three-Dimensional Quasistatic Model for High Brightness Beam
                   Dynamics Simulation},
  journal       = {Phys. Rev. ST Accel. Beams},
  volume        = {9},
  pages         = {044204},
  year          = {2006},
  doi           = {10.1103/PhysRevSTAB.9.044204}}

@mastersthesis{Liemen:2025,
  author        = {Liemen, Alexander},
  title         = {Adaptive Energy Binning in {OPALX}},
  school        = {ETH Zurich},
  year          = {2025},
  url           = {https://github.com/aliemen/HS24-masters-thesis/releases/tag/v2.0-2025.07}}

@article{Prat:2022,
  author        = {Prat, Eduard and Craievich, Paolo and Dijkstal, Philipp 
                   and Di Mitri, Simone and Ferrari, Eugenio and Lucas, Thomas G. 
                   and Malyzhenkov, Alexander and Perosa, Giovanni 
                   and Reiche, Sven and Schietinger, Thomas},
  title         = {Energy spread blowup by intrabeam scattering 
                   and microbunching at the {SwissFEL} injector},
  journal       = {Phys. Rev. Accel. Beams},
  volume        = {25},
  number        = {10},
  pages         = {104401},
  year          = {2022},
  doi           = {10.1103/PhysRevAccelBeams.25.104401}}

@book{Hockney:1988,
  author        = {Hockney, R. W. and Eastwood, J. W.},
  title         = {{Computer Simulation Using Particles}},
  publisher     = {{CRC} Press},
  address       = {Boca Raton},
  year          = {2021},
  doi           = {10.1201/9780367806934}}

@misc{wyssling_boundary_2021,
  author        = {Wyssling, Tim and Adelmann, Andreas},
  title         = {On {Boundary} {Conditions} in the sub-mesh interaction 
                   of the {Particle}-{Particle}-{Particle}-{Mesh} {Algorithm}},
  year          = {2021},
  eprint        = {2103.02046},
  archivePrefix = {arXiv},
  primaryClass  = {physics.comp-ph},
  url           = {https://arxiv.org/abs/2103.02046}}

@manual{Floettmann:ASTRA,
  author       = {Fl\"ottmann, Klaus},
  title        = {{ASTRA}: A Space Charge Tracking Algorithm},
  organization = {DESY},
  address      = {Hamburg, Germany},
  year         = {2017},
  note         = {User manual, version 3.2},
  url          = {https://www.desy.de/~mpyflo/Astra_manual/Astra-Manual_V3.2.pdf}}

@book{Hairer:2006,
  author        = {Hairer, Ernst and Lubich, Christian and Wanner, Gerhard},
  title         = {Geometric {Numerical} {Integration}: {Structure-Preserving 
                   Algorithms for Ordinary Differential Equations}},
  series        = {Springer {Series} in {Computational} {Mathematics}},
  volume        = {31},
  edition       = {2nd},
  publisher     = {Springer},
  address       = {Berlin, Heidelberg},
  year          = {2006},
  doi           = {10.1007/3-540-30666-8}}

@article{Rodrigues:1840,
  author        = {Rodrigues, Olinde}, 
  title         = {Des lois géométriques qui régissent les déplacements d'un système 
                   solide dans l'espace, et de la variation des coordonnées provenant 
                   de ces déplacements considérés indépendamment des causes qui peuvent 
                   les produire},
  journal       = {Journal de Mathématiques Pures et Appliquées},
  volume        = {5},
  pages         = {380--440},
  year          = {1840},
  url           = {https://numdam.org/item/JMPA_1840_1_5__380_0/}}

@article{Dai:2015,
  author        = {Dai, Jian S.},
  title         = {Euler–{Rodrigues} formula variations, quaternion conjugation and intrinsic connections},
  journal       = {Mechanism and Machine Theory},
  volume        = {92},
  pages         = {144--152},
  year          = {2015},
  doi           = {10.1016/j.mechmachtheory.2015.03.004}}

@article{Zenitani:2018,
  author        = {Zenitani, Seiji and Umeda, Takayuki},
  title         = {On the {Boris} solver in particle-in-cell simulation},
  journal       = {Physics of Plasmas},
  volume        = {25},
  number        = {11},
  pages         = {112110},
  year          = {2018},
  doi           = {10.1063/1.5051077}}

@inproceedings{Mitchell:2015,
  author        = {Mitchell, Chad and Qiang, Ji},
  title         = {A {Parallel} {Particle}-{Particle}, {Particle}-{Mesh} 
                   {Solver} for {Studying} {Coulomb} {Collisions} in the 
                   {Code} {IMPACT}-{T}},
  booktitle     = {Proc. 6th International Particle Accelerator 
                   Conference (IPAC'15)},
  pages         = {593--595},
  publisher     = {JACoW, Geneva, Switzerland},
  year          = {2015},
  doi           = {10.18429/JACoW-IPAC2015-MOPMA024}}

@misc{Huebl:2026,
  author        = {Huebl, Axel},
  title         = {{ImpactX} {Benchmarks}},
  year          = {2026},
  howpublished  = {\url{https://github.com/BLAST-ImpactX/impactx-benchmarks}},
  note          = {{Accessed: 16 October 2026}}}

@inproceedings{Piwinski:1974,
  author        = "Piwinski, A.",
  title         = "{Intra-beam-Scattering}",
  booktitle     = "{9th International Conference on High-Energy Accelerators}",
  reportNumber  = "DESY-H1-73-3",
  pages         = "405--409",
  year          = "1974",
  venue         = {SLAC, Stanford, CA, USA}}

@article{Bjorken:1983,
  author        = "Bjorken, James D and Mtingwa, S K",
  title         = "{Intrabeam scattering}",
  journal       = "Part. Accel.",
  reportNumber  = "FERMILAB-PUB-82-47-THY",
  volume        = "13",
  pages         = "115--143",
  year          = "1983",
  url           = "https://cds.cern.ch/record/140304"}

@article{Bane:2002,
  author 		  = {K. L. F. Bane and H. Hayano and K. Kubo and T. Naito 
                   and T. Okugi and J. Urakawa},
  title 		  = "{Intrabeam scattering analysis of measurements at {KEK}'s 
                    Accelerator Test Facility damping ring}",
  journal 	  = {Physical Review Special Topics - Accelerators and Beams},
  year 		  = "2002",
  month 		  = {Aug},
  publisher 	  = {American Physical Society ({APS})},
  volume 		  = {5},
  number 		  = {8},
  url 		  = {https://doi.org/10.1103\%2Fphysrevstab.5.084403},
  doi 		  = {10.1103/physrevstab.5.084403}}

@article{Lucas:2026,
  author        = {Lucas, Thomas G. and Craievich, Paolo and 
                   Prat, Eduard and Reiche, Sven and Gjonaj, Erion},
  title         = {Accurate modelling of intrabeam scattering and
                   its impact on photoinjectors for free-electron lasers},
  journal       = {Scientific Reports},
  volume        = {16},
  pages         = {2629},
  year          = {2026},
  doi           = {10.1038/s41598-026-36558-3}}

@article{DiMitri:2020,
  author        = {Di Mitri, S and Perosa, G and Brynes, A and Setija, I 
                   and Spampinati, S and Williams, P H and Wolski, A 
                   and Allaria, E and Brussaard, S and Giannessi, L 
                   and others},
  title         = {Experimental evidence of intrabeam scattering in a 
                   free-electron laser driver},
  journal       = {New Journal of Physics},
  volume        = {22},
  number        = {8},
  pages         = {083053},
  year          = {2020},
  doi           = {10.1088/1367-2630/aba572}}

@article{jegerlehner_muon_2009,
  author        = {Jegerlehner, Fred and Nyffeler, Andreas},
  title         = {The muon g-2},
  journal       = {Physics Reports},
  volume        = {477},
  number        = {1-3},
  pages         = {1--110},
  year          = {2009},
  doi           = {10.1016/j.physrep.2009.04.003}}

@article{adelmann_compact_2025,
  author        = {Adelmann, A. and Bainbridge, A. R. and Bailey, I. 
                   and Baldini, A. and Basnet, S. and Berger, N. 
                   and Bianco, L. and Calzolaio, C. and Caminada, L. 
                   and Cavoto, G. and others},
  title         = {A compact frozen-spin trap for the search for the 
                   electric dipole moment of the muon},
  volume        = {85},
  number        = {6},
  journal       = {The European Physical Journal C},
  year          = {2025},
  pages         = {622},
  doi           = {10.1140/epjc/s10052-025-14295-7}}

@article{farley_new_2004,
  author        = {Farley, F. J. M. and Jungmann, K. and Miller, J. P. 
                   and Morse, W. M. and Orlov, Y. F. and Roberts, B. L. 
                   and Semertzidis, Y. K. and Silenko, A. and Stephenson, E. J.},
  title         = {New {Method} of {Measuring} {Electric} {Dipole} {Moments} 
                   in {Storage} {Rings}},
  journal       = {Physical Review Letters},
  volume        = {93},
  number        = {5},
  pages         = {052001},
  year          = {2004},
  doi           = {10.1103/PhysRevLett.93.052001}}

@inproceedings{roberts_g4beamline_2007,
  author        = {Roberts, Thomas J. and Kaplan, Daniel M.},
  title         = {G4beamline simulation program for matter-dominated beamlines},
  booktitle     = {2007 {IEEE} {Particle} {Accelerator} {Conference} ({PAC})},
  pages         = {3468--3470},
  publisher     = {IEEE},
  year          = {2007},
  doi           = {10.1109/PAC.2007.4440461}}

@article{wien_untersuchungen_1898,
  author        = {Wien, Willy},
  title         = {Untersuchungen {\"u}ber die electrische
                   {Entladung} in verd{\"u}nnten {Gasen}},
  journal       = {Annalen der Physik},
  volume        = {301},
  number        = {6},
  pages         = {440--452},
  year          = {1898},
  doi           = {10.1002/andp.18983010618}}

@article{plies_wien_2011,
  author        = {Plies, E. and Marianowski, K. and Ohnweiler, T.},
  title         = {The {Wien} filter: {History}, fundamentals and modern applications},
  journal       = {Nuclear Instruments and Methods in Physics Research Section A},
  volume        = {645},
  number        = {1},
  pages         = {7--11},
  year          = {2011},
  doi           = {10.1016/j.nima.2010.12.215}}

@incollection{tsuno_chapter_2013,
  author        = {Tsuno, Katsushige and Ioanoviciu, Damaschin},
  title         = {Application of {Wien} {Filters} to {Electrons}},
  booktitle     = {Advances in {Imaging} and {Electron} {Physics}},
  volume        = {176},
  pages         = {177--212},
  publisher     = {Elsevier},
  year          = {2013},
  doi           = {10.1016/B978-0-12-408142-0.00006-7}}

@article{agostinelli_geant4_2003,
  author        = {Agostinelli, S. and others},
  title         = {{GEANT4} - {A} {Simulation} {Toolkit}},
  journal       = {Nucl. Instrum. Meth. A},
  volume        = {506},
  number        = {3},
  pages         = {250--303},
  year          = {2003},
  doi           = {10.1016/S0168-9002(03)01368-8}}

@inproceedings{UriotTraceWin:2015,
  author        = {Uriot, Didier and Pichoff, Nicolas},
  title         = {{Status of TraceWin Code}},
  booktitle     = {Proc. 6th International Particle Accelerator Conference (IPAC'15)},
  pages         = {92--94},
  publisher     = {JACoW},
  year          = {2015},
  doi           = {10.18429/JACoW-IPAC2015-MOPWA008}}

@article{prokscha_new_2008,
  author        = {Prokscha, T. and Morenzoni, E. and Deiters, K. 
                   and Foroughi, F. and George, D. and Kobler, R. 
                   and Suter, A. and Vrankovic, V.},
  title         = {The new {$\mu\mathrm{E4}$} beam at {PSI}: 
                   {A} hybrid-type large acceptance channel for the 
                   generation of a high intensity surface-muon beam},
  journal       = {Nucl. Instrum. Methods Phys. Res. A},
  volume        = {595},
  number        = {2},
  pages         = {317--331},
  year          = {2008},
  doi           = {10.1016/j.nima.2008.07.081}}

@article{Maxson:2013,
  author        = {Maxson, J. M. and Bazarov, I. V. and Wan, W. 
                  and Padmore, H. A. and Coleman-Smith, C. E.},
  title         = {Fundamental photoemission brightness limit from disorder induced heating},
  journal       = {New Journal of Physics},
  volume        = {15},
  pages         = {103024},
  year          = {2013},
  doi           = {10.1088/1367-2630/15/10/103024}}

\end{document}